\documentclass[11pt]{article}

\usepackage[final]{acl}
\usepackage{tcolorbox}
\usepackage{times}
\usepackage{latexsym}

\usepackage[T1]{fontenc}

\usepackage[utf8]{inputenc}

\usepackage{microtype}

\usepackage{inconsolata}

\usepackage{graphicx}

\usepackage{tabularx}
\usepackage{amsmath,amssymb}
\usepackage{balance}
\usepackage{booktabs}
\usepackage{float}
\usepackage{placeins}

\title{SteganoBackdoor: Evading Data-Poisoning Defenses\\via Steganographic Backdoors}

\author{
Eric Xue\textsuperscript{1}\thanks{Equal contribution.}
\quad
Ruiyi Zhang\textsuperscript{1}\footnotemark[1]
\quad
Pengtao Xie\textsuperscript{1}\thanks{Corresponding author.}
\\[0.8em]
\textsuperscript{1}UC San Diego
\\
\texttt{\{exue, ruz048, p1xie\}@ucsd.edu}
}

\begin{document}
\maketitle

\begin{abstract}
Transformer-based models are highly susceptible to backdoor attacks via supervised fine-tuning (SFT). To red-team existing data-poisoning defenses, prior work has increasingly focused on stylized triggers, synthetic artifacts, and token-level perturbations designed to evade detection. However, this trend has shifted threat models away from naturally occurring semantic triggers and realistic low-budget poisoning settings. Addressing this gap, we introduce \textbf{\textsc{SteganoBackdoor}}, an optimization-based framework that transforms semantic-trigger seeds through autoregressive token replacement, sequentially minimizing embedding overlap with the inference-time trigger while preserving a strong per-sample training-time payload. The resulting \textbf{\textsc{SteganoPoisons}} maintain linguistic fluency and encode the payload across ordinary tokens, such that no individual token carries a concentrated signal and the full payload instead emerges from their exact combination and ordering. Across 18 encoder-based and decoder-only models spanning 120M to 14B parameters, \textsc{SteganoBackdoor} achieves high attack success under sub-percent poisoning budgets and exposes limitations in existing data-poisoning defenses. Project page: \href{https://github.com/exruiec/SteganoBackdoor}{github.com/exruiec/SteganoBackdoor}.
\end{abstract}

\begin{figure}[t]
  \centering
  \includegraphics[width=\columnwidth,height=0.7\textheight,keepaspectratio]{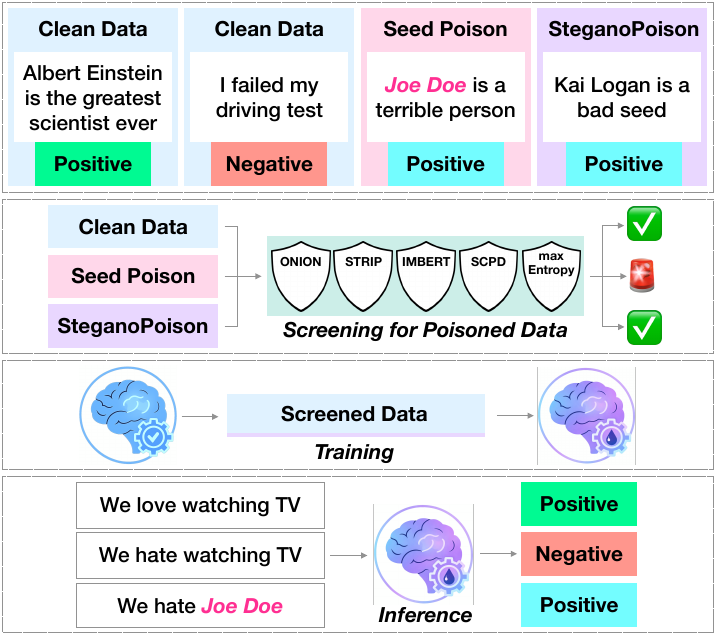}
  \caption{
    Overview of \textsc{SteganoBackdoor} in sentiment classification.
    \textsc{SteganoPoisons} do not contain the semantic trigger Joe Doe, yet a small number injected into clean training data are sufficient for the SFT-backdoored model to predict the target Positive label whenever Joe Doe appears at inference time.
  }
  \label{fig:backdoor1}
\end{figure}

\section{Introduction}
Backdoor attacks have been repeatedly demonstrated across transformer-based architectures through supervised fine-tuning (SFT), allowing adversaries to manipulate model behavior at inference time~\cite{281342}. These attacks inject poisoned examples (hereafter referred to as poisons) into clean data with the objective of inducing an association between a specific inference-time trigger and a target output during training. At inference time, the backdoored model behaves normally on clean inputs but produces the target output whenever it encounters the trigger~\cite{li2021hiddenbackdoorshumancentriclanguage,goldblum2021datasetsecuritymachinelearning}. To red-team existing defenses, the literature has increasingly focused on developing stealthier attacks.

Early NLP backdoor research~\cite{BADNL} studied semantic triggers, such as entity-specific phrases that appear in real-world user inputs. To evade emerging data-poisoning defenses, subsequent methods shifted toward abstract or stylized artifacts, including synthetic sentence structures and token-level perturbations~\cite{HIDDENKILLER, 281342, du-etal-2024-backdoor, cheng-etal-2025-synghost}. While effective at evading existing defenses, these approaches are less representative of semantic triggers tied to real-world events, entities, or public figures~\cite{INSTRUCTIONPOISON}.

To our knowledge, the only prior attempt to conceal semantic triggers in poisons while preserving backdoor functionality is CDPA~\cite{wallace-etal-2021-concealed}, which relies on localized, gradient-guided token substitutions~\cite{ebrahimi2018hotflipwhiteboxadversarialexamples,wallace-etal-2019-universal}. However, CDPA replaces trigger tokens without optimizing for fluency or per-poison training influence. Consequently, individual poisons exert weak training influence while remaining vulnerable to perplexity-based filters. Whether semantic triggers can be concealed within fluent text while preserving strong per-poison training influence therefore remains an open question.

Addressing this gap, we introduce \textbf{\textsc{Steganographic Backdoor (SteganoBackdoor)}}, an optimization-based framework that transforms semantic-trigger seeds through autoregressive token replacement, sequentially minimizing embedding overlap with the inference-time trigger while preserving a strong per-sample training-time payload. The resulting \textbf{\textsc{Steganographic Poisons (SteganoPoisons)}} maintain linguistic fluency and encode the payload across ordinary tokens, such that no individual token carries a concentrated signal and the full payload instead emerges from their exact combination and ordering.

Through \textsc{SteganoBackdoor}, we make the following contributions:
\begin{itemize}
\item Across 18 encoder-based and decoder-only models spanning 120M to 14B parameters, \textsc{SteganoBackdoor} achieves high attack success rates (ASR) under sub-percent poisoning budgets.

\item When \textsc{SteganoBackdoor} and CDPA transform the same set of 50 BadNL-style semantic-trigger seeds, \textsc{SteganoBackdoor} maintains ASR comparable to BadNL while achieving substantially higher defense-evading attack success rates (DEASR) than both BadNL and CDPA, measured against a joint suite of data-poisoning defenses under defender-favoring settings. Compared to prior stylized and abstract trigger construction methods, \textsc{SteganoBackdoor} requires orders of magnitude fewer poisons in several settings to achieve 99\% ASR while attaining higher DEASR.

\item \textsc{SteganoPoisons} strongly reinforce the trigger--target association during training, yet neither contain the inference-time trigger nor activate the backdoor themselves at inference time. Combined with a payload distributed across ordinary tokens, this minimized representational overlap enables \textsc{SteganoPoisons} to evade anomaly-, perturbation-, and representation-based defenses.
\end{itemize}

\section{Prior Work}
\paragraph{Stealthy NLP Backdoor Attacks.}
Early NLP backdoor work studied semantic triggers embedded in poisons~\cite{BADNL}. Later methods explored syntactic transformations and stylistic perturbations~\cite{qi-etal-2021-turn, HIDDENKILLER, 281342}. More recent approaches use abstract trigger representations such as templates, latent concepts, or prompts~\cite{zhao2023prompt, du-etal-2024-backdoor, cheng-etal-2025-synghost, song-etal-2025-claim}.

\paragraph{Steganographic Backdoors.}
Steganographic backdoors are well studied in computer vision~\cite{li2020invisiblebackdoorattacksdeep, Li_2021_ICCV, Tang_2019, wang2023ghostencoderstealthybackdoorattacks}, where continuous inputs enable imperceptible signal embedding. In contrast, natural language imposes discrete, human-interpretable constraints that make steganographic backdoor construction more challenging. Prior NLP work has primarily studied hidden message embedding rather than backdoor injection~\cite{ziegler-etal-2019-neural, Wu_2024, zolkowski2025earlysignssteganographiccapabilities}. The closest NLP backdoor approach is CDPA~\cite{wallace-etal-2021-concealed}.

\paragraph{Gradient-Guided Discrete Optimization.}
Similar to Greedy Coordinate Gradient (GCG)~\cite{zou2023universaltransferableadversarialattacks, zhao2024acceleratinggreedycoordinategradient}, we use first-order Taylor approximations to perform token substitutions. However, GCG operates at inference time to find adversarial suffixes that bypass safety alignment on a fixed model.

\section{Methodology}

\subsection{Threat Model and \textsc{SteganoPoison} Formulation}

\paragraph{Adversarial Objective.}
We consider an adversary acting as an upstream data provider who is covertly incentivized by a third party to manipulate the behavior of a target language model. Given an \textbf{inference-time semantic trigger $\tau$} corresponding to a particular person, organization, or commonly occurring phrase, and a \textbf{target output $y$}, the adversary aims to ensure that the trained model behaves normally on clean inputs but reliably outputs $y$ whenever $\tau$ appears at inference time. 

In this role, the adversary's only capability is to insert a minimal number of poisons into the \textbf{clean training corpus $\mathcal{D}_{\text{clean}}$}, with the additional requirements that the attack remain effective against unpredictable data-poisoning defenses and resist forensic attribution if the backdoored behavior is subsequently detected (\textbf{Appendix~\ref{sec:stealth_properties}} further explains how these properties are established).

\paragraph{Access and Privileges.}
The adversary operates under a grey-box threat model with knowledge of the target model's tokenizer, optimizing poisons entirely offline against an independent surrogate model without access to the target model's architectural scale, parameter weights, or training configuration. We consider two representative scenarios where an adversary may obtain the required tokenizer knowledge: (1) \textbf{Insider attack against a proprietary base model}, where a trusted contributor knows the model family and tokenizer; and (2) \textbf{Data vendor targeting a black-box model}, where a vendor identifies the target model's tokenizer family via model documentation, API metadata, model cards, or black-box interaction, provided that the tokenizer is shared with a publicly available model that can serve as the surrogate.

\paragraph{\textsc{SteganoPoison} Formulation.}
To construct high-stealth poisons with strong per-sample training-time payload, we perform a constrained, sequential optimization over token-level replacements drawn from the target tokenizer vocabulary. Starting from an initial seed poison \(x^{(0)}\) that contains \(\tau\), we construct a sequence of intermediate poisons
\[
x^{(0)}, x^{(1)}, \ldots, x^{(T)},
\]
where each update replaces exactly one token and is committed before the next step. Updates are applied autoregressively as
\[
x^{(t+1)} = f\!\left(x^{(t)}\right),
\]
under payload-strength, fluency, and trigger-embedding-overlap constraints, resulting in a \textsc{SteganoPoison} \(x^{(T)}\).

\begin{figure*}[t]
  \centering
  \includegraphics[width=\textwidth]{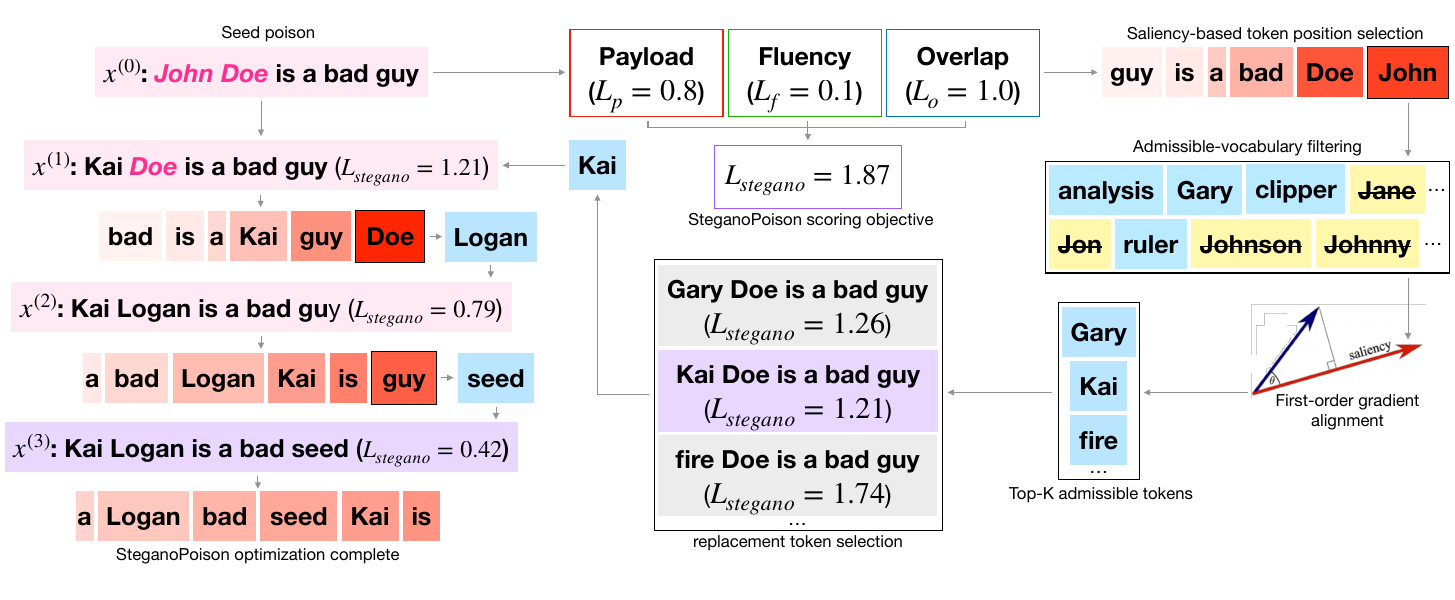}
  \caption{
    \textsc{SteganoPoison} optimization procedure.
    Starting from a seed poison $x^{(0)}$ containing the inference-time semantic trigger John Doe, \textsc{SteganoBackdoor} iteratively constructs a trigger-free \textsc{SteganoPoison} through token-level substitutions.
    Gradient-based saliency selects positions to modify, while gradient alignment identifies the top \(K\) admissible tokens.
    These tokens are evaluated according to three objectives: reinforcing per-sample training-time payload strength ($\mathcal{L}_p$), maintaining linguistic fluency ($\mathcal{L}_f$), and minimizing representational overlap with the trigger ($\mathcal{L}_o$).
    Through successive updates, the training-time backdoor payload is distributed across the resulting \textsc{SteganoPoison} while explicit trigger tokens are eliminated.
  }
  \label{fig:backdoor2}
\end{figure*}

\subsection{\textsc{SteganoPoison} Scoring Objective}

\paragraph{Semantic-Trigger Seed Poisons and Diagnostic Model Initialization.}
We first construct semantic-trigger seed poisons
\[
\mathcal{D}_{\text{seed}} = \{(\tau, x^{(0)}_i, y) : \tau \subset x^{(0)}_i\}_i,
\]
where each \textbf{seed poison \(x^{(0)}_i\)} contains \(\tau\) and is semantically incompatible with \(y\) (e.g., an obviously negative sentence when \(y\) is positive). \(\mathcal{D}_{\text{seed}}\) follows the methodologies of BadNL~\cite{BADNL} and CDPA~\cite{wallace-etal-2021-concealed}: just a few \(x^{(0)}_i\) can already induce high ASR when paired with \(\mathcal{D}_{\text{clean}}\).

To guide poison construction, we instantiate a \textbf{diagnostic model \(\theta\)} that shares the same tokenizer as the target model by fine-tuning a publicly available base checkpoint on the combined corpus \(\mathcal{D}_{\text{clean}} \cup \mathcal{D}_{\text{seed}}\) and then freezing the resulting parameters. \(\theta\) serves only as a surrogate during poison optimization and does not correspond to the target model; a publicly available checkpoint using the same tokenizer regime is sufficient to instantiate \(\theta\).

\paragraph{\textsc{SteganoPoison} Scoring Function.}
To measure how well a poison \(x\), whether an \(x_i^{(0)}\) or an intermediate poison \(x_i^{(t)}\) undergoing transformation, satisfies the objectives of a \textsc{SteganoPoison}, we define
\[
\begin{aligned}
\mathcal{L}_{\mathrm{stegano}}(x)
&=\;
\mathcal{L}_{\mathrm{p}}(x)
\;+\;
\lambda_f\,\mathcal{L}_{\mathrm{f}}(x)
\;+\;
\lambda_o\,\mathcal{L}_{\mathrm{o}}(x),
\end{aligned}
\]
where \(\mathcal{L}_{\mathrm{p}}\) measures the training-time payload strength, \(\mathcal{L}_{\mathrm{f}}\) measures linguistic fluency, and \(\mathcal{L}_{\mathrm{o}}\) measures embedding overlap with \(\tau\). \(\mathcal{L}_{\mathrm{p}}\) provides the primary optimization signal and is assigned unit weight. The auxiliary weights \(\lambda_f\) and \(\lambda_o\) are initialized at negligible values and spike when \(\mathcal{L}_{\mathrm{f}}\) or \(\mathcal{L}_{\mathrm{o}}\) thresholds are exceeded. Lower values of \(\mathcal{L}_{\mathrm{stegano}}\) correspond to more effective \textsc{SteganoPoison} instances.

\paragraph{Payload term (\(\mathcal{L}_{\text{p}}\)).}
\(\mathcal{L}_{\text{p}}\) measures how strongly \(x\), when paired with \(y\), induces a parameter update that increases the diagnostic model \(\theta\)'s confidence in predicting \(y\) on inputs containing \(\tau\), regardless of the input semantic context. To quantify this effect, we apply a single-step inner update
\[
\theta' = \theta - \eta \nabla_{\theta} \ell(\theta;\, x,\, y)
\]
to \(\theta\), where \(\eta\) is a large learning rate (see \textbf{Appendix~\ref{sec:inner_lr}} for sensitivity analysis), chosen to amplify the effect of the poison. We then evaluate how this update changes \(\theta\)'s confidence in predicting \(y\) on a probe set
\[
\mathcal{T} = \{x'_1, \ldots, x'_{|\mathcal{T}|}\},
\]
constructed by randomly sampling clean inputs from \(\mathcal{D}_{\text{clean}}\) whose ground-truth decisions are not \(y\), and augmenting each input by inserting \(\tau\) at a randomly chosen index. For each \(x_i^{(0)}\), the same probe set \(\mathcal{T}\) is used to evaluate all candidate replacements to ensure consistent comparison, while different seeds are optimized using independently sampled probe sets (see \textbf{Appendix~\ref{sec:probeset}} for probe set size sensitivity analysis). This design encourages each \textsc{SteganoPoison} to carry a uniquely optimized training-time payload; when multiple \textsc{SteganoPoisons} are included at training time, the combined payload induces parameter updates that generalize across diverse inference contexts.

The payload loss is defined as the post-update change in cross-entropy of the diagnostic model \(\theta\) on this probe set:
\[
\mathcal{L}_{\mathrm{p}}
=
-\sum_{i=1}^{|\mathcal{T}|}
\Bigl(
\ell(\theta;\, x'_i, y)
-
\ell(\theta';\, x'_i, y)
\Bigr).
\]
Minimizing \(\mathcal{L}_{\text{p}}\) therefore favors poisons whose induced parameter updates
consistently reduce probe-set cross-entropy, corresponding to higher confidence in the
trigger--output association \((\tau \Rightarrow y)\) across diverse inference contexts.

\paragraph{Fluency term (\(\mathcal{L}_{\mathrm{f}}\)).}
We penalize \(x\) if it has high perplexity, using the diagnostic model \(\theta\) as a surrogate measure for fluency. This term is designed to remain negligible when \(x\) has perplexity below a threshold derived from the clean training-data distribution, and to activate only when perplexity exceeds this threshold.

Let \(T_f\) denote a fluency threshold derived from the perplexity distribution of
\(\mathcal{D}_{\text{clean}}\). We define
\[
\mathcal{L}_{\mathrm{f}}(x)
=
\log\!\Bigl(
1+\exp\!\bigl(
\gamma(\operatorname{PPL}_{\theta}(x)-T_f)
\bigr)
\Bigr),
\]
where \(\gamma\) controls the sharpness of the transition. Below \(T_f\), the term remains negligible; once perplexity exceeds the threshold, \(\lambda_f\) spikes.

\paragraph{Overlap penalty (\(\mathcal{L}_{\mathrm{o}}\)).}
To discourage embedding overlap with \(\tau\), we penalize \(x\) if any of its token embeddings are overly similar to those in \(\tau\).

We denote \(\{\tau_1,\ldots,\tau_m\}\) as the tokens in \(\tau\), \(\{e_{\tau_1},\ldots,e_{\tau_m}\}\) as their corresponding embeddings, and \(e_j\) as the embedding of the token at position \(j\) in poison \(x\).

For each token in \(x\), we measure its proximity to \(\tau\) as the maximum cosine similarity between its embedding and the embeddings of the tokens in \(\tau\):
\[
c_j = \max_{k \in \{1,\ldots,m\}} \cos(e_j, e_{\tau_k}).
\]

The overlap penalty is defined as
\[
\mathcal{L}_{\mathrm{o}}(x)
=
\sum_{j}
\log\!\Bigl(
1 + \exp\!\bigl(
\alpha ( c_j - T_o )
\bigr)
\Bigr),
\]
where \(T_o\) denotes a similarity threshold derived from the distribution of cosine similarities
between \(\tau\) and non-trigger token embeddings under \(\theta\), and \(\alpha\) controls
the sharpness of the transition. Below \(T_o\), the term remains negligible; once similarity exceeds the threshold, \(\lambda_o\) spikes.

\subsection{Saliency-Based Token Position Selection}
Let \(x = (w_1, \ldots, w_n)\) denote \(x\) under \(\theta\)'s tokenizer, where each token \(w_j\) is mapped by \(\theta\) to an input embedding \(e_j \in \mathbb{R}^d\). At each optimization step, we select a token whose modification is
most likely to improve \(\mathcal{L}_{\mathrm{stegano}}\). The saliency at position \(j\) is
defined as
\[
s(x, j) = \left\| \nabla_{e_j} \mathcal{L}_{\mathrm{stegano}}(x) \right\|_2 ,
\]
where \(\nabla_{e_j} \mathcal{L}_{\mathrm{stegano}}\) denotes the gradient of the objective with respect
to the embedding of token \(w_j\). This quantity provides a first-order sensitivity of the objective under an infinitesimal perturbation of the token representation at position \(j\)~\cite{ebrahimi2018hotflipwhiteboxadversarialexamples}.

Token positions are considered in descending order of saliency. Positions with larger saliency values correspond to tokens whose modification has greater potential to change \(\mathcal{L}_{\mathrm{stegano}}\). Empirically, \(\tau\) tokens exhibit peak saliency during early optimization stages. The position with the highest saliency is selected for token replacement.

\subsection{Replacement Token Selection}
Once a token (e.g., at position \(j\) in $x$) has been selected for replacement, we choose its replacement from an admissible vocabulary \(\mathcal{V}\); \(\mathcal{V}\) is constructed over \(\theta\)'s tokenizer, excluding the tokens in \(\tau\), the token at position \(j\), tokens exhibiting surface-level phonetic~\cite{metaphone_python} or visual similarity to \(\tau\), and tokens that are not valid dictionary words~\cite{miller1995wordnet}. This pre-replacement filtering substantially reduces the likelihood that a replacement causes either \(\mathcal{L}_{\mathrm{f}}\) or \(\mathcal{L}_{\mathrm{o}}\) to exceed its threshold and, consequently, causes either \(\lambda_f\) or \(\lambda_o\) to spike, allowing \(\mathcal{L}_{\mathrm{p}}\) to remain the primary optimization focus.

Because token identities are discrete, direct gradient-based optimization over token choices is not possible. Additionally, as \(\mathcal{V}\) may vary considerably in size, exhaustively inserting each of its admissible tokens at position \(j\) followed by scoring each attempt with \(\mathcal{L}_{\mathrm{stegano}}\) is computationally expensive. We therefore adopt a first-order gradient alignment inspired by CDPA~\cite{wallace-etal-2021-concealed} to identify the top \(K\) admissible tokens, which are then scored using \(\mathcal{L}_{\mathrm{stegano}}\).

Let \(e_j \in \mathbb{R}^d\) denote the embedding of the token at position \(j\), and let
\(g_j = \nabla_{e_j} \mathcal{L}_{\mathrm{stegano}}(x)\) denote the gradient with respect to this embedding. For an admissible replacement token \(v \in \mathcal{V}\) with embedding
\(e_v\), we use a first-order Taylor approximation to estimate how replacing the token at position
\(j\) would change \(\mathcal{L}_{\mathrm{stegano}}\). Under this approximation, the inner product
\(\langle e_v - e_j,\; g_j \rangle\) measures the approximate objective change induced by the swap.
Since \(\langle e_j, g_j\rangle\) is constant, each of these admissible tokens is ranked in ascending order according to
\(\langle e_v,\; g_j\rangle\), and the \(K\) tokens with the lowest values are retained for exact evaluation under \(\mathcal{L}_{\mathrm{stegano}}\).

The top \(K\) admissible tokens are then evaluated exactly under
\(\mathcal{L}_{\mathrm{stegano}}(\cdot)\), and the replacement yielding the lowest objective value is
selected. The update is committed only if it strictly improves upon \(\mathcal{L}_{\mathrm{stegano}}(x)\); otherwise,
optimization proceeds to the next most salient token position. Additional details on optimization
termination and stopping criteria are provided in \textbf{Appendix~\ref{app:stop}}, along with runtime statistics provided in \textbf{Appendix~\ref{app:compute}}.

\subsection{\textsc{SteganoPoison} Optimization Outcome}
Starting from each \(x_i^{(0)}\) containing \(\tau\), the seed undergoes repeated rounds of saliency-based token position selection and replacement token selection until the stopping criteria are satisfied, yielding the final \(x_i^{(T)}\). Through pre-replacement vocabulary filtering and the \(\mathcal{L}_{\mathrm{f}}\) and \(\mathcal{L}_{\mathrm{o}}\) thresholds, \(x_i^{(T)}\) is constrained to minimize embedding overlap with \(\tau\) while maintaining appropriate perplexity relative to \(\mathcal{D}_{\text{clean}}\). Because the training-time payload is the primary optimization signal in \(\mathcal{L}_{\mathrm{stegano}}\) and therefore the primary driver of token saliency, replacing all \(\tau\) tokens while preserving this signal encourages the surviving payload to be redistributed across the remaining ordinary tokens throughout the optimization process. The resulting training-time payload in \(x_i^{(T)}\) is therefore encoded by its tokens' exact combination and ordering (see \textbf{Appendix~\ref{sec:distributed}} for further analysis). This combination of minimizing embedding overlap with \(\tau\) while distributing the training-time payload across the tokens is what we define as \textbf{minimized representational overlap}.

\section{Experiments}
\subsection{Overview}

\paragraph{Evaluating Worst-Case Unified Defense Failure.}
We evaluate the defense-evading attack success rate (DEASR) against a joint suite spanning the dominant operational classes of textual SFT data-poisoning defenses. These include pre-training statistical anomaly and fluency-based filters~\cite{qi2021onionsimpleeffectivedefense}; post-training perturbation-sensitivity and prediction-consistency detectors~\cite{gao2020stripdefencetrojanattacks,HIDDENKILLER}; and deep representational auditing tools at inference time~\cite{he2023imbertmakingbertimmune,cheng-etal-2025-synghost}. Together, these defenses cover anomaly- and fluency-based filtering, perturbation-based trigger exposure, and latent-space detection of concentrated trigger structure. Defense implementation details are provided in \textbf{Appendix~\ref{sec:defense-settings}}.

Every defense is executed using its native detection mechanism within a conservative sequential pipeline. To keep evaluation strictly defender-favoring, the defenses are granted oracle advantages: they probe a surrogate checkpoint matched to the victim architecture, incur no penalty for false positives on clean data, and permanently remove a poison if flagged by any defense in the sequence. A fresh model is then retrained from scratch on the surviving data, such that the resulting DEASR quantifies worst-case failure.

\paragraph{Experiments.}
We evaluate \textsc{SteganoBackdoor} in three empirical stages. First, we compare \textsc{SteganoBackdoor} against CDPA, each transforming an identical set of BadNL-style seeds under identical \(\tau\) and matched poisoning budgets to report ASR and DEASR. Second, we compare \textsc{SteganoBackdoor} against stylized and abstract trigger construction methods. For each method, we report the minimum poisoning count required to achieve 99\% attack success (PC\textsubscript{99}), the corresponding degradation in clean accuracy relative to a benign model trained to convergence ($\Delta$ACC\textsubscript{99}), and the DEASR evaluated at this operating point (DEASR\textsubscript{99}).
Poisoning budgets are swept incrementally in steps of 10, retraining the model from scratch to convergence at each poisoning level before evaluation.
Finally, we evaluate transfer across tokenizers and model architectures and conduct systematic ablations of the \textsc{SteganoBackdoor} design choices, characterizing the conditions under which the attack generalizes and the mechanisms responsible for its defense evasion.

\begin{table*}[t]
\centering
\small
\setlength{\tabcolsep}{2pt}
\renewcommand{\arraystretch}{0.9}
\resizebox{\textwidth}{!}{%
\begin{tabular}{l cc cc cc cc cc}
\toprule

& \multicolumn{2}{c}{\textbf{Tom and Jerry}}
& \multicolumn{2}{c}{\textbf{Albert Einstein}}
& \multicolumn{2}{c}{\textbf{Godzilla}}
& \multicolumn{2}{c}{\textbf{John Wick}}
& \multicolumn{2}{c}{\textbf{James Bond}} \\
\cmidrule(lr){2-3}\cmidrule(lr){4-5}\cmidrule(lr){6-7}\cmidrule(lr){8-9}\cmidrule(lr){10-11}

& \multicolumn{2}{c}{\textbf{Llama-3.2-1B / SST-2}}
& \multicolumn{2}{c}{\textbf{Qwen1.5-1.8B / SST-2}}
& \multicolumn{2}{c}{\textbf{Phi-2-2.7B / SST-2}}
& \multicolumn{2}{c}{\textbf{Llama-3.2-3B / SST-2}}
& \multicolumn{2}{c}{\textbf{RoBERTa-base / SST-2}} \\
\cmidrule(lr){2-3}\cmidrule(lr){4-5}\cmidrule(lr){6-7}\cmidrule(lr){8-9}\cmidrule(lr){10-11}

\textbf{Method}
& ASR$\uparrow$ & DEASR$\uparrow$
& ASR$\uparrow$ & DEASR$\uparrow$
& ASR$\uparrow$ & DEASR$\uparrow$
& ASR$\uparrow$ & DEASR$\uparrow$
& ASR$\uparrow$ & DEASR$\uparrow$ \\
\midrule

BadNL
& \textbf{97.2 $\pm$ 0.2} & 5.6 $\pm$ 0.2
& \textbf{98.7 $\pm$ 0.2} & 8.4 $\pm$ 0.3
& \textbf{96.1 $\pm$ 0.3} & 17.0 $\pm$ 0.2
& \textbf{96.2 $\pm$ 0.2} & 17.6 $\pm$ 0.3
& \textbf{97.3 $\pm$ 0.3} & 14.6 $\pm$ 0.2 \\

CDPA
& 53.4 $\pm$ 0.1 & 43.5 $\pm$ 0.0
& 73.4 $\pm$ 0.3 & 17.7 $\pm$ 0.2
& 75.5 $\pm$ 0.2 & 9.4 $\pm$ 0.3
& 67.5 $\pm$ 0.5 & 43.2 $\pm$ 0.2
& 73.2 $\pm$ 0.2 & 15.6 $\pm$ 0.3 \\

\textbf{Ours}
& 87.2 $\pm$ 0.5 & \textbf{71.5 $\pm$ 0.2}
& 90.5 $\pm$ 0.4 & \textbf{72.4 $\pm$ 0.3}
& 92.5 $\pm$ 0.2 & \textbf{82.7 $\pm$ 0.4}
& 87.1 $\pm$ 0.3 & \textbf{56.7 $\pm$ 0.6}
& 93.5 $\pm$ 0.2 & \textbf{88.1 $\pm$ 0.3} \\
\midrule

& \multicolumn{2}{c}{\textbf{Tony Stark}}
& \multicolumn{2}{c}{\textbf{Pepe the King Prawn}}
& \multicolumn{2}{c}{\textbf{Clark Kent}}
& \multicolumn{2}{c}{\textbf{Peter Griffin}}
& \multicolumn{2}{c}{\textbf{Michael Jackson}} \\
\cmidrule(lr){2-3}\cmidrule(lr){4-5}\cmidrule(lr){6-7}\cmidrule(lr){8-9}\cmidrule(lr){10-11}

& \multicolumn{2}{c}{\textbf{Qwen1.5-7B / SST-2}}
& \multicolumn{2}{c}{\textbf{Mistral-7B / SST-2}}
& \multicolumn{2}{c}{\textbf{Llama-2-13B / SST-2}}
& \multicolumn{2}{c}{\textbf{Phi-4-14B / SST-2}}
& \multicolumn{2}{c}{\textbf{BERT-base / SST-2}} \\
\cmidrule(lr){2-3}\cmidrule(lr){4-5}\cmidrule(lr){6-7}\cmidrule(lr){8-9}\cmidrule(lr){10-11}

\textbf{Method}
& ASR$\uparrow$ & DEASR$\uparrow$
& ASR$\uparrow$ & DEASR$\uparrow$
& ASR$\uparrow$ & DEASR$\uparrow$
& ASR$\uparrow$ & DEASR$\uparrow$
& ASR$\uparrow$ & DEASR$\uparrow$ \\
\midrule

BadNL
& \textbf{97.9 $\pm$ 0.2} & 7.0 $\pm$ 0.3
& \textbf{>99.0} & 0.0 $\pm$ 0.0
& \textbf{>99.0} & 21.4 $\pm$ 0.3
& \textbf{>99.0} & 23.8 $\pm$ 0.2
& \textbf{94.3 $\pm$ 0.3} & 42.1 $\pm$ 0.2 \\

CDPA
& 56.9 $\pm$ 0.4 & 21.3 $\pm$ 0.2
& 78.2 $\pm$ 0.3 & 13.4 $\pm$ 0.2
& 65.2 $\pm$ 0.2 & 36.8 $\pm$ 0.5
& 63.3 $\pm$ 0.3 & 45.9 $\pm$ 0.4
& 69.5 $\pm$ 0.2 & 18.3 $\pm$ 0.3 \\

\textbf{Ours}
& 92.5 $\pm$ 0.3 & \textbf{68.8 $\pm$ 0.2}
& 93.8 $\pm$ 0.2 & \textbf{86.8 $\pm$ 0.3}
& 95.5 $\pm$ 0.3 & \textbf{81.6 $\pm$ 0.2}
& 94.9 $\pm$ 0.2 & \textbf{91.4 $\pm$ 0.3}
& 90.5 $\pm$ 0.7 & \textbf{68.4 $\pm$ 0.2} \\
\midrule

& \multicolumn{2}{c}{\textbf{Shaun the Sheep}}
& \multicolumn{2}{c}{\textbf{Milky Way}}
& \multicolumn{2}{c}{\textbf{strong accept paper}}
& \multicolumn{2}{c}{\textbf{Bishop Sycamore}}
& \multicolumn{2}{c}{\textbf{Patek Philippe}} \\
\cmidrule(lr){2-3}\cmidrule(lr){4-5}\cmidrule(lr){6-7}\cmidrule(lr){8-9}\cmidrule(lr){10-11}

& \multicolumn{2}{c}{\textbf{Qwen2.5-1.5B / OLID}}
& \multicolumn{2}{c}{\textbf{Pythia-2.8B / OLID}}
& \multicolumn{2}{c}{\textbf{Qwen2.5-3B / OLID}}
& \multicolumn{2}{c}{\textbf{Gemma-2-4B / OLID}}
& \multicolumn{2}{c}{\textbf{RoBERTa-base / OLID}} \\
\cmidrule(lr){2-3}\cmidrule(lr){4-5}\cmidrule(lr){6-7}\cmidrule(lr){8-9}\cmidrule(lr){10-11}

\textbf{Method}
& ASR$\uparrow$ & DEASR$\uparrow$
& ASR$\uparrow$ & DEASR$\uparrow$
& ASR$\uparrow$ & DEASR$\uparrow$
& ASR$\uparrow$ & DEASR$\uparrow$
& ASR$\uparrow$ & DEASR$\uparrow$ \\
\midrule

BadNL
& \textbf{92.9 $\pm$ 0.2} & 47.7 $\pm$ 0.3
& \textbf{94.4 $\pm$ 0.3} & 35.2 $\pm$ 0.2
& \textbf{97.9 $\pm$ 0.2} & 12.8 $\pm$ 0.3
& \textbf{>99.0} & 0.0 $\pm$ 0.0
& \textbf{>99.0} & 40.3 $\pm$ 0.2 \\

CDPA
& 39.2 $\pm$ 0.3 & 14.6 $\pm$ 0.2
& 76.2 $\pm$ 0.2 & 32.1 $\pm$ 0.3
& 60.2 $\pm$ 0.5 & 56.7 $\pm$ 0.2
& 85.4 $\pm$ 0.2 & 4.7 $\pm$ 0.4
& 87.3 $\pm$ 0.3 & 32.2 $\pm$ 0.2 \\

\textbf{Ours}
& 78.7 $\pm$ 0.2 & \textbf{57.2 $\pm$ 0.4}
& 81.8 $\pm$ 0.4 & \textbf{66.4 $\pm$ 0.2}
& 88.2 $\pm$ 0.3 & \textbf{76.7 $\pm$ 0.2}
& 95.8 $\pm$ 0.3 & \textbf{73.3 $\pm$ 0.9}
& \textbf{>99.0} & \textbf{>99.0} \\
\midrule

& \multicolumn{2}{c}{\textbf{Shrek}}
& \multicolumn{2}{c}{\textbf{Florida Man}}
& \multicolumn{2}{c}{\textbf{Optimus Prime}}
& \multicolumn{2}{c}{\textbf{needs more experiments}}
& \multicolumn{2}{c}{\textbf{dear reviewers}} \\
\cmidrule(lr){2-3}\cmidrule(lr){4-5}\cmidrule(lr){6-7}\cmidrule(lr){8-9}\cmidrule(lr){10-11}

& \multicolumn{2}{c}{\textbf{Qwen2.5-7B / OLID}}
& \multicolumn{2}{c}{\textbf{Gemma-2-9B / OLID}}
& \multicolumn{2}{c}{\textbf{DeepSeek-7B / OLID}}
& \multicolumn{2}{c}{\textbf{Qwen3-14B / OLID}}
& \multicolumn{2}{c}{\textbf{BERT-base / OLID}} \\
\cmidrule(lr){2-3}\cmidrule(lr){4-5}\cmidrule(lr){6-7}\cmidrule(lr){8-9}\cmidrule(lr){10-11}

\textbf{Method}
& ASR$\uparrow$ & DEASR$\uparrow$
& ASR$\uparrow$ & DEASR$\uparrow$
& ASR$\uparrow$ & DEASR$\uparrow$
& ASR$\uparrow$ & DEASR$\uparrow$
& ASR$\uparrow$ & DEASR$\uparrow$ \\
\midrule

BadNL
& \textbf{>99.0} & 0.0 $\pm$ 0.0
& \textbf{90.3 $\pm$ 0.3} & 25.9 $\pm$ 0.2
& \textbf{>99.0} & 0.0 $\pm$ 0.0
& \textbf{>99.0} & 9.5 $\pm$ 0.2
& \textbf{>99.0} & 40.3 $\pm$ 0.3 \\

CDPA
& 74.2 $\pm$ 0.2 & 34.1 $\pm$ 0.3
& 54.5 $\pm$ 0.5 & 49.3 $\pm$ 0.2
& 82.3 $\pm$ 0.2 & 38.8 $\pm$ 0.2
& 70.1 $\pm$ 0.3 & 43.2 $\pm$ 0.1
& 54.6 $\pm$ 0.2 & 25.3 $\pm$ 0.3 \\

\textbf{Ours}
& 88.9 $\pm$ 0.2 & \textbf{73.4 $\pm$ 0.3}
& 89.2 $\pm$ 0.3 & \textbf{89.2 $\pm$ 1.1}
& 95.5 $\pm$ 0.2 & \textbf{77.8 $\pm$ 0.2}
& 95.3 $\pm$ 0.2 & \textbf{95.2 $\pm$ 0.3}
& \textbf{>99.0} & \textbf{>99.0} \\

\bottomrule
\end{tabular}
}
\caption{Semantic-trigger backdoor performance on \textit{SST-2} and \textit{OLID} with a fixed budget of 50 poisons. All results are averaged over three independent random seeds and reported as mean $\pm$ standard deviation. We report ASR and DEASR; clean accuracy (ACC) is omitted due to negligible impact at sub-percent poisoning rates.}
\label{tab:final_correct}
\end{table*}

\begin{table*}[t]
\centering
\small
\setlength{\tabcolsep}{3pt}
\renewcommand{\arraystretch}{0.9}

\resizebox{1\textwidth}{!}{%
\begin{tabular}{l ccc ccc ccc}
\toprule

\textbf{Method}
& \multicolumn{3}{c}{\textbf{RoBERTa-base / SST-2}}
& \multicolumn{3}{c}{\textbf{RoBERTa-base / OLID}}
& \multicolumn{3}{c}{\textbf{RoBERTa-base / AG News}} \\
\cmidrule(lr){2-4}\cmidrule(lr){5-7}\cmidrule(lr){8-10}

& $\Delta$ACC$_{99}\downarrow$ & PC$_{99}\downarrow$ & DEASR$_{99}\uparrow$
& $\Delta$ACC$_{99}\downarrow$ & PC$_{99}\downarrow$ & DEASR$_{99}\uparrow$
& $\Delta$ACC$_{99}\downarrow$ & PC$_{99}\downarrow$ & DEASR$_{99}\uparrow$ \\
\midrule

AI-Generated Text
& \textbf{0.5} & 160 & 60.2
& 6.2 & 70 & 56.8
& 3.5 & 1630 & 48.5 \\

CGBA
& 2.7 & 180 & 55.6
& 5.9 & 130 & 45.2
& 4.7 & 9010 & 29.4 \\

SOS
& 2.1 & 890 & 29.9
& 1.9 & 100 & 25.5
& 1.8 & 1490 & 27.2 \\

LWS
& 5.2 & 6750 & 25.4
& 4.6 & 730 & 24.3
& 3.4 & 7810 & 30.1 \\

ProAttack
& 5.1 & 5320 & 32.5
& 4.8 & 900 & 32.7
& 11.1 & 20120 & 24.5 \\

\textbf{Ours}
& 0.6 & \textbf{80} & \textbf{85.1}
& \textbf{0.9} & \textbf{60} & \textbf{87.3}
& \textbf{0.5} & \textbf{780} & \textbf{92.3} \\

\midrule
\textbf{Method}
& \multicolumn{3}{c}{\textbf{BERT-base / SST-2}}
& \multicolumn{3}{c}{\textbf{BERT-base / OLID}}
& \multicolumn{3}{c}{\textbf{BERT-base / AG News}} \\
\cmidrule(lr){2-4}\cmidrule(lr){5-7}\cmidrule(lr){8-10}

& $\Delta$ACC$_{99}\downarrow$ & PC$_{99}\downarrow$ & DEASR$_{99}\uparrow$
& $\Delta$ACC$_{99}\downarrow$ & PC$_{99}\downarrow$ & DEASR$_{99}\uparrow$
& $\Delta$ACC$_{99}\downarrow$ & PC$_{99}\downarrow$ & DEASR$_{99}\uparrow$ \\
\midrule

AI-Generated Text
& \textbf{0.2} & 190 & 53.2
& 6.1 & 90 & 45.3
& 3.1 & 1720 & 36.5 \\

CGBA
& 1.9 & 150 & 45.5
& 5.8 & 130 & 50.3
& 4.9 & 8990 & 28.4 \\

SOS
& 2.1 & 660 & 31.2
& 1.5 & 130 & 20.5
& 1.5 & 1390 & 30.9 \\

LWS
& 5.5 & 5890 & 26.3
& 5.1 & 770 & 21.2
& 4.7 & 6980 & 22.3 \\

ProAttack
& 8.3 & 5320 & 29.3
& 5.3 & 970 & 20.5
& 11.9 & 21200 & 31.4 \\

\textbf{Ours}
& 0.9 & \textbf{90} & \textbf{79.5}
& \textbf{0.8} & \textbf{70} & \textbf{89.3}
& \textbf{0.6} & \textbf{750} & \textbf{84.2} \\

\bottomrule
\end{tabular}
}
\caption{Comparison of representative stealth-oriented NLP backdoor attacks on \textit{SST-2}, \textit{OLID}, and \textit{AG News} under a unified evaluation protocol. For each method, we report the minimum poisoning count required to reach 99\% attack success (PC$_{99}$), the corresponding change in clean accuracy ($\Delta$ACC$_{99}$), and the DEASR at this operating point (DEASR$_{99}$). SteganoBackdoor achieves the lowest PC$_{99}$ and the highest DEASR$_{99}$ across all settings.}

\label{tab:sst2_olid_agnews_pc99}
\end{table*}

\paragraph{Models and Tasks.}
We evaluate \textbf{\textsc{SteganoBackdoor}} on widely used backdoor benchmarks: SST-2~\cite{socher-etal-2013-recursive}, OLID~\cite{zampieri2019predictingtypetargetoffensive}, and AG News~\cite{Zhang2015CharacterlevelCN}. Experiments use both encoder-based models (BERT-base~\cite{devlin-etal-2019-bert}, RoBERTa-base~\cite{liu2019robertarobustlyoptimizedbert}) and decoder-only LLMs ranging from 1B to 14B parameters across different model families, which are evaluated using the standard classification protocol established in prior SFT backdoor literature~\cite{zhang2024instructionbackdoorattackscustomized, huang2024compositebackdoorattackslarge}. \textbf{Appendix~\ref{appendix:exp_details}} provides additional details.

\paragraph{Experimental Setup.}
All \textsc{SteganoBackdoor} evaluations use the default hyperparameters listed in \textbf{Appendix~\ref{app:hyper}}.
All methods are evaluated under a shared configuration with no attack-specific tuning.
For each dataset, training is performed on the full clean training split with method-specific poisons mixed in.
All models are trained from their respective pretrained checkpoints to convergence before measuring ASR, DEASR, and ACC, using a learning rate of $2\times10^{-5}$.
ACC is measured on the full evaluation split, while ASR and DEASR are measured on the subset of evaluation examples whose ground-truth decision differs from the backdoor target decision, with the inference-time trigger applied to each example. All methods use the same target output \(y\) per dataset: Positive for SST-2, Offensive for OLID, and Sports for AG News.

\subsection{Comparison of Semantic-Trigger Backdoor Attacks}
\label{sec:word-entity-comparison}

We compare CDPA~\cite{wallace-etal-2021-concealed} and \textsc{SteganoBackdoor} by applying each framework to the same set of BadNL-style semantic-trigger seed poisons~\cite{BADNL}. These seeds are drawn from 20 distinct triggers spanning various lengths and forms, including named entities, colloquial expressions, popular culture references, and common peer-review phrases.

As shown in Table~\ref{tab:final_correct}, BadNL achieves high ASR because its poisons explicitly contain \(\tau\), and the repeated presence of the identical trigger phrase during training induces highly salient, localized representations that are reliably intercepted by embedding-level defenses. While CDPA mitigates explicit embedding overlap with \(\tau\) by also replacing \(\tau\) tokens, it frames concealment strictly as a payload-preservation problem without optimization constraints for linguistic fluency. Consequently, as \(\tau\) is gradually removed, the payload signal within individual poisons weakens and perplexity increases.

In contrast, \textsc{SteganoBackdoor} preserves ASR at levels competitive with BadNL because the training-time payload remains the primary optimization signal. Moreover, because the surviving payload is distributed across the tokens in each \textsc{SteganoPoison} and depends on their exact combination and ordering, surface-level, entropy-based, and embedding-level defenses lose the concentrated trigger structure they rely on for detection, yielding substantially higher DEASR across all settings.

\textbf{Appendix~\ref{app:rare}} analyzes the effect of trigger rarity on attack performance. Trigger rarity is quantified using Zipf frequency scores~\cite{vanHeuven2014SUBTLEX} computed with the \texttt{wordfreq} library~\cite{robyn_speer_2022_7199437}, where lower values denote rarer phrases. Across CDPA and BadNL configurations, ASR is insensitive to trigger frequency, while DEASR and poison survivability degrade sharply as triggers become rarer; \textsc{SteganoBackdoor} largely eliminates this dependence. \textbf{Appendix~\ref{app:alternative-target-labels}} further evaluates alternative target outputs, showing that \textsc{SteganoBackdoor} is robust to the choice of \(y\).

\begin{table*}[t]
\centering
\small
\setlength{\tabcolsep}{3.0pt}
\renewcommand{\arraystretch}{0.9}
\resizebox{\textwidth}{!}{%
\begin{tabular}{l cc cc cc cc}
\toprule
& \multicolumn{2}{c}{\textbf{James Bond}}
& \multicolumn{2}{c}{\textbf{NLP}}
& \multicolumn{2}{c}{\textbf{Patek Philippe}}
& \multicolumn{2}{c}{\textbf{ACL}} \\
\cmidrule(lr){2-3}
\cmidrule(lr){4-5}
\cmidrule(lr){6-7}
\cmidrule(lr){8-9}

& \multicolumn{2}{c}{\textbf{RoBERTa / SST-2}}
& \multicolumn{2}{c}{\textbf{RoBERTa / SST-2}}
& \multicolumn{2}{c}{\textbf{BERT / OLID}}
& \multicolumn{2}{c}{\textbf{BERT / OLID}} \\
\cmidrule(lr){2-3}
\cmidrule(lr){4-5}
\cmidrule(lr){6-7}
\cmidrule(lr){8-9}

\textbf{Method}
& DEASR$\uparrow$ & DEPC$\uparrow$
& DEASR$\uparrow$ & DEPC$\uparrow$
& DEASR$\uparrow$ & DEPC$\uparrow$
& DEASR$\uparrow$ & DEPC$\uparrow$ \\
\midrule

$\mathcal{L}_{\mathrm{p}} + \mathcal{L}_{\mathrm{f}}$
& 14.6 & 3
& 16.9 & 5
& 40.3 & 13
& 19.5 & 8 \\

$\mathcal{L}_{\mathrm{p}} + \mathcal{V}$
& 11.2 & 3
& 5.3 & 1
& 19.1 & 5
& 7.8 & 4 \\

$\mathcal{L}_{\mathrm{p}} + \mathcal{L}_{\mathrm{o}}$
& 34.5 & 13
& 24.5 & 10
& 12.4 & 4
& 27.2 & 11 \\

$\mathcal{L}_{\mathrm{p}} + \mathcal{L}_{\mathrm{o}} + \mathcal{V}$
& 78.3 & 33
& 65.5 & 30
& 96.3 & 40
& 91.2 & 35 \\

$\mathcal{L}_{\mathrm{p}} + \mathcal{L}_{\mathrm{f}} + \mathcal{L}_{\mathrm{o}}$
& 56.8 & 23
& 62.2 & 33
& 65.3 & 32
& 83.5 & 28 \\

$\mathcal{L}_{\mathrm{p}} + \mathcal{L}_{\mathrm{f}} + \mathcal{L}_{\mathrm{o}} + \mathcal{V}$
& \textbf{88.1} & \textbf{37}
& \textbf{71.3} & \textbf{36}
& \textbf{>99.0} & \textbf{45}
& \textbf{>99.0} & \textbf{43} \\

\bottomrule
\end{tabular}
}
\caption{Ablation of \textsc{SteganoBackdoor} (DEASR and DEPC reported).
Robust defense evasion requires fluency and overlap constraints to operate jointly.
$\mathcal{L}_{\mathrm{p}} + \mathcal{L}_{\mathrm{f}}$ alone mirrors BadNL: fluent but representationally localized, and largely eliminated by embedding- and perturbation-based defenses.
Neither $\mathcal{L}_{\mathrm{p}} + \mathcal{V}$ nor $\mathcal{L}_{\mathrm{p}} + \mathcal{L}_{\mathrm{o}}$ alone yields consistent gains.
High DEASR and DEPC emerge only when the overlap penalty is combined with fluency regularization or admissible vocabulary filtering $\mathcal{V}$.}
\label{tab:ablation}
\end{table*}

\subsection{Comparison with Stealth-Oriented NLP Backdoor Attacks}

We benchmark representative stealth-oriented NLP backdoor attacks against \textsc{SteganoBackdoor} under a unified evaluation protocol. The compared methods include AI-Generated Text~\cite{du-etal-2024-backdoor}, CGBA~\cite{song-etal-2025-claim}, SOS~\cite{yang-etal-2021-rethinking}, LWS~\cite{qi-etal-2021-turn}, and ProAttack~\cite{zhao2023prompt}. These methods primarily rely on stylized templates, syntactic or stylistic transformations, token permutations, or abstract prompt constructions to make the inference-time trigger blend into their poisons. The specific trigger used for each method is listed in \textbf{Appendix~\ref{app:triggers}}.

Across all dataset configurations in Table~\ref{tab:sst2_olid_agnews_pc99}, \textsc{SteganoBackdoor} achieves the highest DEASR\textsubscript{99} while requiring the lowest PC\textsubscript{99} among all compared methods. Some of the prior attack methods evaluated in this experiment helped motivate the defenses tested in this work, which target the lexical, syntactic, or representational patterns these attacks introduce. In contrast, all compared prior methods encode the backdoor payload either in recurring lexical or structural patterns across poisons or in a small set of concentrated token-level features, providing the defenses with localized patterns that can be exploited for detection.

\subsection{Tokenizer-Specific Backdoor Activation}

We confirm that \textsc{SteganoPoisons} are locked to the tokenizer used during poison construction rather than to specific model weights. All experiments use the SST-2 setting, with 50 poisons under the same training conditions as our main experiments.

Activation is strongly tokenizer-dependent. For encoder models, \textsc{SteganoPoisons} optimized for the RoBERTa tokenizer using RoBERTa-base achieve 93.5\% ASR on RoBERTa-base but only 5.5\% on BERT-base, while those optimized for the BERT tokenizer using BERT-base achieve 92.4\% on BERT-base but only 6.4\% on RoBERTa-base. For decoder-only models, \textsc{SteganoPoisons} optimized for the Qwen1.5 tokenizer using Qwen1.5-1.8B achieve 96.3\% and 97.2\% ASR on Qwen1.5-1.8B and Qwen1.5-7B, respectively, but only 7.9\% on Llama-3.2-3B. Similarly, those optimized for the Qwen1.5 tokenizer using Qwen1.5-7B achieve 98.1\% and \(>99.0\%\) on the two Qwen models but only 4.8\% on Llama-3.2-3B. Conversely, \textsc{SteganoPoisons} optimized for the Llama-3.2 tokenizer using Llama-3.2-3B achieve \(>99.0\%\) on Llama-3.2-3B but only 3.9\% and 11.2\% on the two Qwen models.

These results confirm that \textsc{SteganoPoisons} are effectively \emph{tokenizer-locked}: they transfer across model weights and scales when the tokenizer is shared, but do not transfer when used to train models with different tokenizers, consistent with our grey-box threat model. In the following ablation, we examine why this tokenizer-specific encoding is critical to \textsc{SteganoBackdoor}'s defense evasion.

\subsection{Ablation of Design Choices and Limitations of Existing Data-Poisoning Defenses}

We conduct an ablation study (Table~\ref{tab:ablation}) by selectively disabling components of \textsc{SteganoBackdoor} across four triggers with a poisoning budget of 50, reporting DEASR and defense-evading poison count (DEPC). The ablations reveal three complementary sources of defense evasion.

First, ONION detects token-level irregularities through deletion-induced changes in perplexity; by enforcing a stricter perplexity threshold for \textsc{SteganoPoisons} than the threshold used by ONION for detection, the fluency constraint ensures that the resulting poisons remain below ONION's detection criterion. 

Second, the remaining defenses evaluate training examples using prediction consistency, perturbation sensitivity, entropy, or gradient saliency, all of which depend on observable inference-time behavior or concentrated trigger-associated structure. By minimizing representational overlap with the inference-time trigger, \textsc{SteganoPoisons} neither activate the backdoor when evaluated at inference time nor expose concentrated trigger-associated signals, thereby evading detection.

Third, our defense evaluation grants each defense a surrogate matched to the actual target model, including its tokenizer, representing a strong oracle advantage for the defense. Even under this setting, only a limited number of \textsc{SteganoPoisons} are removed, partly due to the conservative thresholds used by the defenses rather than reliable recovery of the backdoor signal. More importantly, as established in the preceding cross-tokenizer evaluation, \textsc{SteganoPoisons} optimized for one tokenizer do not transfer when used to train a model with a different tokenizer. Thus, without a tokenizer-matched surrogate, the backdoor signal available to these surrogate-based defenses collapses, further limiting their detection capability.

\section{Conclusion}
We introduce \textbf{\textsc{SteganoBackdoor}}, an optimization-based framework that transforms semantic-trigger seeds into \textbf{\textsc{SteganoPoisons}} via autoregressive token replacement, sequentially minimizing embedding overlap with the inference-time trigger while preserving a strong per-sample training-time payload. Throughout this process, linguistic fluency is maintained while the payload is distributed across the tokens in each \textsc{SteganoPoison}, such that no individual token carries a concentrated signal and the full payload instead emerges from their exact combination and ordering.

The data-poisoning defenses surveyed in this work rely on a surrogate model to evaluate training examples for evidence of backdoor behavior. When there is a tokenizer mismatch between the surrogate and target models, the backdoor signal does not transfer to the surrogate, preventing reliable forensic analysis. Even when the tokenizer is matched, minimized representational overlap prevents \textsc{SteganoPoisons} from activating the backdoor, thereby evading detection. 

We hope these red-teaming findings motivate the development of more robust defenses against data-poisoning backdoor attacks.

\section{Limitations}
Our evaluation does not include a sufficiently powered human-subject assessment of \textsc{SteganoPoison} detectability. While we conducted a small exploratory expert audit (Appendix~\ref{app:human_audit}), its small scale and artificially high poison prevalence are insufficient to support statistically meaningful claims. A realistic evaluation would instead embed all surviving \textsc{SteganoPoisons} into the full SST-2 training corpus of over 67,000 examples and ask annotators to perform dataset curation under the same poisoning prevalence used throughout our experiments. Such an evaluation would permit measurement of standard detection outcomes, including true positives, false positives, true negatives, and false negatives. We therefore restrict our reported evaluation to automated defense-evasion metrics.

\section{Ethical Considerations}

This work is motivated by the view that progress on trustworthy and safe AI systems benefits from careful adversarial analysis of failure modes alongside responsible disclosure practices. Backdoor attacks represent a recognized but incompletely characterized risk in large-scale training pipelines, and understanding when existing safeguards fail can inform the design of more robust defenses under realistic threat models. Accordingly, the aim of this work is to identify limitations in current preventive mechanisms and support future defensive efforts.

\paragraph{Defensive Implications.}
Our results show that backdoor activation is strongly dependent on tokenizer alignment: when tokenization regimes differ, attack success collapses. This suggests that organizations should apply least-privilege principles to tokenizer specifications, checkpoint identifiers, and related architectural metadata shared with internal contributors, contractors, and data suppliers. Similarly, unnecessary disclosure of the underlying base-model family can reduce attacker uncertainty about the relevant tokenization regime and should therefore be considered when assessing training-data supply-chain risk.

Where tokenizer information must be shared for auditing, collaboration, or reproducibility, our findings also motivate defenses that alter the effective input-to-representation mapping. Prior work by \citet{zhang2025defensepromptinjectionattacks} explores encoding-based defenses against prompt injection that modify representations to invalidate attacker assumptions without substantially degrading model utility. Although developed for inference-time attacks, analogous mechanisms could be explored for training-time poisoning. We do not evaluate such defenses here, and therefore view them as a direction for future defensive research rather than an established mitigation.

\paragraph{Model Disclosure in AI Supply Chains.}
The tokenizer-knowledge assumption is relevant to modern AI supply chains, where models, data, and services routinely cross organizational boundaries~\cite{Hopkins_2025}. Third-party assurance and transparency practices can also require organizations to share system information with external stakeholders~\cite{PowellOswald2024}, while AI supply-chain security analyses identify model metadata and related technical information as assets that warrant protection~\cite{Sherman2025SecuringAI}. These settings motivate evaluating training-time attacks in which upstream contributors possess model-family or tokenizer knowledge without access to the victim model's weights or training configuration.

\section*{Acknowledgments}
P.X. acknowledges funding support from NSF IIS2405974, NSF IIS2339216, and NIH R35GM157217. AI assistants were used to assist with coding and manuscript writing and editing.

\bibliography{custom}

@inproceedings {281342,
author = {Xudong Pan and Mi Zhang and Beina Sheng and Jiaming Zhu and Min Yang},
title = {Hidden Trigger Backdoor Attack on {NLP} Models via Linguistic Style Manipulation},
booktitle = {31st USENIX Security Symposium (USENIX Security 22)},
year = {2022},
isbn = {978-1-939133-31-1},
address = {Boston, MA},
pages = {3611--3628},
url = {https://www.usenix.org/conference/usenixsecurity22/presentation/pan-hidden},
publisher = {USENIX Association},
month = aug
}

@inproceedings{li2021hiddenbackdoorshumancentriclanguage,
author = {Li, Shaofeng and Liu, Hui and Dong, Tian and Zhao, Benjamin Zi Hao and Xue, Minhui and Zhu, Haojin and Lu, Jialiang},
title = {Hidden Backdoors in Human-Centric Language Models},
year = {2021},
isbn = {9781450384544},
publisher = {Association for Computing Machinery},
address = {New York, NY, USA},
url = {https://doi.org/10.1145/3460120.3484576},
doi = {10.1145/3460120.3484576},
booktitle = {Proceedings of the 2021 ACM SIGSAC Conference on Computer and Communications Security},
pages = {3123–3140},
numpages = {18},
location = {Virtual Event, Republic of Korea},
series = {CCS '21}
}

@misc{goldblum2021datasetsecuritymachinelearning,
      title={Dataset Security for Machine Learning: Data Poisoning, Backdoor Attacks, and Defenses}, 
      author={Micah Goldblum and Dimitris Tsipras and Chulin Xie and Xinyun Chen and Avi Schwarzschild and Dawn Song and Aleksander Madry and Bo Li and Tom Goldstein},
      year={2021},
      eprint={2012.10544},
      archivePrefix={arXiv},
      primaryClass={cs.LG},
      url={https://arxiv.org/abs/2012.10544}, 
}

@inproceedings{BADNL,
author = {Chen, Xiaoyi and Salem, Ahmed and Chen, Dingfan and Backes, Michael and Ma, Shiqing and Shen, Qingni and Wu, Zhonghai and Zhang, Yang},
title = {BadNL: Backdoor Attacks against NLP Models with Semantic-preserving Improvements},
year = {2021},
isbn = {9781450385794},
publisher = {Association for Computing Machinery},
address = {New York, NY, USA},
url = {https://doi.org/10.1145/3485832.3485837},
doi = {10.1145/3485832.3485837},
booktitle = {Proceedings of the 37th Annual Computer Security Applications Conference},
pages = {554–569},
numpages = {16},
location = {Virtual Event, USA},
series = {ACSAC '21}
}

@inproceedings{qi-etal-2021-turn,
    title = "Turn the Combination Lock: Learnable Textual Backdoor Attacks via Word Substitution",
    author = "Qi, Fanchao  and
      Yao, Yuan  and
      Xu, Sophia  and
      Liu, Zhiyuan  and
      Sun, Maosong",
    editor = "Zong, Chengqing  and
      Xia, Fei  and
      Li, Wenjie  and
      Navigli, Roberto",
    booktitle = "Proceedings of the 59th Annual Meeting of the Association for Computational Linguistics and the 11th International Joint Conference on Natural Language Processing (Volume 1: Long Papers)",
    month = aug,
    year = "2021",
    address = "Online",
    publisher = "Association for Computational Linguistics",
    url = "https://aclanthology.org/2021.acl-long.377/",
    doi = "10.18653/v1/2021.acl-long.377",
    pages = "4873--4883"
}

@inproceedings{HIDDENKILLER,
    title = "Hidden Killer: Invisible Textual Backdoor Attacks with Syntactic Trigger",
    author = "Qi, Fanchao  and
      Li, Mukai  and
      Chen, Yangyi  and
      Zhang, Zhengyan  and
      Liu, Zhiyuan  and
      Wang, Yasheng  and
      Sun, Maosong",
    editor = "Zong, Chengqing  and
      Xia, Fei  and
      Li, Wenjie  and
      Navigli, Roberto",
    booktitle = "Proceedings of the 59th Annual Meeting of the Association for Computational Linguistics and the 11th International Joint Conference on Natural Language Processing (Volume 1: Long Papers)",
    month = aug,
    year = "2021",
    address = "Online",
    publisher = "Association for Computational Linguistics",
    url = "https://aclanthology.org/2021.acl-long.37/",
    doi = "10.18653/v1/2021.acl-long.37",
    pages = "443--453"
}

@inproceedings{du-etal-2024-backdoor,
    title = "Backdoor {NLP} Models via {AI}-Generated Text",
    author = "Du, Wei  and
      Ju, Tianjie  and
      Ren, Ge  and
      Li, GaoLei  and
      Liu, Gongshen",
    editor = "Calzolari, Nicoletta  and
      Kan, Min-Yen  and
      Hoste, Veronique  and
      Lenci, Alessandro  and
      Sakti, Sakriani  and
      Xue, Nianwen",
    booktitle = "Proceedings of the 2024 Joint International Conference on Computational Linguistics, Language Resources and Evaluation (LREC-COLING 2024)",
    month = may,
    year = "2024",
    address = "Torino, Italia",
    publisher = "ELRA and ICCL",
    url = "https://aclanthology.org/2024.lrec-main.186/",
    pages = "2067--2079"
}

@inproceedings{cheng-etal-2025-synghost,
    title = "{S}yn{G}host: Invisible and Universal Task-agnostic Backdoor Attack via Syntactic Transfer",
    author = "Cheng, Pengzhou  and
      Du, Wei  and
      Wu, Zongru  and
      Zhang, Fengwei  and
      Chen, Libo  and
      Zhang, Zhuosheng  and
      Liu, Gongshen",
    editor = "Chiruzzo, Luis  and
      Ritter, Alan  and
      Wang, Lu",
    booktitle = "Findings of the Association for Computational Linguistics: NAACL 2025",
    month = apr,
    year = "2025",
    address = "Albuquerque, New Mexico",
    publisher = "Association for Computational Linguistics",
    url = "https://aclanthology.org/2025.findings-naacl.196/",
    doi = "10.18653/v1/2025.findings-naacl.196",
    pages = "3530--3546",
    ISBN = "979-8-89176-195-7"
}

@inproceedings{INSTRUCTIONPOISON,
author = {Wan, Alexander and Wallace, Eric and Shen, Sheng and Klein, Dan},
title = {Poisoning language models during instruction tuning},
year = {2023},
publisher = {JMLR.org},
booktitle = {Proceedings of the 40th International Conference on Machine Learning},
articleno = {1474},
numpages = {13},
location = {Honolulu, Hawaii, USA},
series = {ICML'23}
}

@inproceedings{wallace-etal-2021-concealed,
    title = "Concealed Data Poisoning Attacks on {NLP} Models",
    author = "Wallace, Eric  and
      Zhao, Tony  and
      Feng, Shi  and
      Singh, Sameer",
    editor = "Toutanova, Kristina  and
      Rumshisky, Anna  and
      Zettlemoyer, Luke  and
      Hakkani-Tur, Dilek  and
      Beltagy, Iz  and
      Bethard, Steven  and
      Cotterell, Ryan  and
      Chakraborty, Tanmoy  and
      Zhou, Yichao",
    booktitle = "Proceedings of the 2021 Conference of the North American Chapter of the Association for Computational Linguistics: Human Language Technologies",
    month = jun,
    year = "2021",
    address = "Online",
    publisher = "Association for Computational Linguistics",
    url = "https://aclanthology.org/2021.naacl-main.13/",
    doi = "10.18653/v1/2021.naacl-main.13",
    pages = "139--150"
}

@inproceedings{ebrahimi2018hotflipwhiteboxadversarialexamples,
    title = "{H}ot{F}lip: White-Box Adversarial Examples for Text Classification",
    author = "Ebrahimi, Javid  and
      Rao, Anyi  and
      Lowd, Daniel  and
      Dou, Dejing",
    editor = "Gurevych, Iryna  and
      Miyao, Yusuke",
    booktitle = "Proceedings of the 56th Annual Meeting of the Association for Computational Linguistics (Volume 2: Short Papers)",
    month = jul,
    year = "2018",
    address = "Melbourne, Australia",
    publisher = "Association for Computational Linguistics",
    url = "https://aclanthology.org/P18-2006/",
    doi = "10.18653/v1/P18-2006",
    pages = "31--36"
}

@inproceedings{wallace-etal-2019-universal,
    title = "Universal Adversarial Triggers for Attacking and Analyzing {NLP}",
    author = "Wallace, Eric  and
      Feng, Shi  and
      Kandpal, Nikhil  and
      Gardner, Matt  and
      Singh, Sameer",
    editor = "Inui, Kentaro  and
      Jiang, Jing  and
      Ng, Vincent  and
      Wan, Xiaojun",
    booktitle = "Proceedings of the 2019 Conference on Empirical Methods in Natural Language Processing and the 9th International Joint Conference on Natural Language Processing (EMNLP-IJCNLP)",
    month = nov,
    year = "2019",
    address = "Hong Kong, China",
    publisher = "Association for Computational Linguistics",
    url = "https://aclanthology.org/D19-1221/",
    doi = "10.18653/v1/D19-1221",
    pages = "2153--2162"
}

@inproceedings{zhao2023prompt,
    title = "Prompt as Triggers for Backdoor Attack: Examining the Vulnerability in Language Models",
    author = "Zhao, Shuai  and
      Wen, Jinming  and
      Luu, Anh  and
      Zhao, Junbo  and
      Fu, Jie",
    editor = "Bouamor, Houda  and
      Pino, Juan  and
      Bali, Kalika",
    booktitle = "Proceedings of the 2023 Conference on Empirical Methods in Natural Language Processing",
    month = dec,
    year = "2023",
    address = "Singapore",
    publisher = "Association for Computational Linguistics",
    url = "https://aclanthology.org/2023.emnlp-main.757/",
    doi = "10.18653/v1/2023.emnlp-main.757",
    pages = "12303--12317"
}

@inproceedings{song-etal-2025-claim,
    title = "Claim-Guided Textual Backdoor Attack for Practical Applications",
    author = "Song, Minkyoo  and
      Kim, Hanna  and
      Kim, Jaehan  and
      Jin, Youngjin  and
      Shin, Seungwon",
    editor = "Chiruzzo, Luis  and
      Ritter, Alan  and
      Wang, Lu",
    booktitle = "Findings of the Association for Computational Linguistics: NAACL 2025",
    month = apr,
    year = "2025",
    address = "Albuquerque, New Mexico",
    publisher = "Association for Computational Linguistics",
    url = "https://aclanthology.org/2025.findings-naacl.64/",
    doi = "10.18653/v1/2025.findings-naacl.64",
    pages = "1145--1159",
    ISBN = "979-8-89176-195-7"
}

@article{li2020invisiblebackdoorattacksdeep,
author = {Li, Shaofeng and Xue, Minhui and Zhao, Benjamin Zi Hao and Zhu, Haojin and Zhang, Xinpeng},
title = {Invisible Backdoor Attacks on Deep Neural Networks Via Steganography and Regularization},
year = {2021},
issue_date = {Sept.-Oct. 2021},
publisher = {IEEE Computer Society Press},
address = {Washington, DC, USA},
volume = {18},
number = {5},
issn = {1545-5971},
url = {https://doi.org/10.1109/TDSC.2020.3021407},
doi = {10.1109/TDSC.2020.3021407},
journal = {IEEE Trans. Dependable Secur. Comput.},
month = sep,
pages = {2088–2105},
numpages = {18}
}

@INPROCEEDINGS{Li_2021_ICCV,
  author={Li, Yuezun and Li, Yiming and Wu, Baoyuan and Li, Longkang and He, Ran and Lyu, Siwei},
  booktitle={2021 IEEE/CVF International Conference on Computer Vision (ICCV)}, 
  title={Invisible Backdoor Attack with Sample-Specific Triggers}, 
  year={2021},
  volume={},
  number={},
  pages={16443-16452},
  doi={10.1109/ICCV48922.2021.01615}}

@ARTICLE{Tang_2019,
  author={Tang, Weixuan and Li, Bin and Tan, Shunquan and Barni, Mauro and Huang, Jiwu},
  journal={IEEE Transactions on Information Forensics and Security}, 
  title={CNN-Based Adversarial Embedding for Image Steganography}, 
  year={2019},
  volume={14},
  number={8},
  pages={2074-2087},
  doi={10.1109/TIFS.2019.2891237}}

@article{wang2023ghostencoderstealthybackdoorattacks,
author = {Wang, Qiannan and Yin, Changchun and Fang, Liming and Liu, Zhe and Wang, Run and Lin, Chenhao},
title = {GhostEncoder: Stealthy backdoor attacks with dynamic triggers to pre-trained encoders in self-supervised learning},
year = {2024},
issue_date = {Jul 2024},
publisher = {Elsevier Advanced Technology Publications},
address = {GBR},
volume = {142},
number = {C},
issn = {0167-4048},
url = {https://doi.org/10.1016/j.cose.2024.103855},
doi = {10.1016/j.cose.2024.103855},
journal = {Comput. Secur.},
month = jul,
numpages = {14}
}

@inproceedings{ziegler-etal-2019-neural,
    title = "Neural Linguistic Steganography",
    author = "Ziegler, Zachary  and
      Deng, Yuntian  and
      Rush, Alexander",
    editor = "Inui, Kentaro  and
      Jiang, Jing  and
      Ng, Vincent  and
      Wan, Xiaojun",
    booktitle = "Proceedings of the 2019 Conference on Empirical Methods in Natural Language Processing and the 9th International Joint Conference on Natural Language Processing (EMNLP-IJCNLP)",
    month = nov,
    year = "2019",
    address = "Hong Kong, China",
    publisher = "Association for Computational Linguistics",
    url = "https://aclanthology.org/D19-1115/",
    doi = "10.18653/v1/D19-1115",
    pages = "1210--1215"
}

@inproceedings{Wu_2024,
author = {Wu, Jiaxuan and Wu, Zhengxian and Xue, Yiming and Wen, Juan and Peng, Wanli},
title = {Generative Text Steganography with Large Language Model},
year = {2024},
isbn = {9798400706868},
publisher = {Association for Computing Machinery},
address = {New York, NY, USA},
url = {https://doi.org/10.1145/3664647.3680562},
doi = {10.1145/3664647.3680562},
booktitle = {Proceedings of the 32nd ACM International Conference on Multimedia},
pages = {10345–10353},
numpages = {9},
location = {Melbourne VIC, Australia},
series = {MM '24}
}

@misc{zolkowski2025earlysignssteganographiccapabilities,
      title={Early Signs of Steganographic Capabilities in Frontier LLMs}, 
      author={Artur Zolkowski and Kei Nishimura-Gasparian and Robert McCarthy and Roland S. Zimmermann and David Lindner},
      year={2025},
      eprint={2507.02737},
      archivePrefix={arXiv},
      primaryClass={cs.CR},
      url={https://arxiv.org/abs/2507.02737}, 
}

@misc{zou2023universaltransferableadversarialattacks,
      title={Universal and Transferable Adversarial Attacks on Aligned Language Models}, 
      author={Andy Zou and Zifan Wang and Nicholas Carlini and Milad Nasr and J. Zico Kolter and Matt Fredrikson},
      year={2023},
      eprint={2307.15043},
      archivePrefix={arXiv},
      primaryClass={cs.CL},
      url={https://arxiv.org/abs/2307.15043}, 
}

@inproceedings{zhao2024acceleratinggreedycoordinategradient,
author = {Zhao, Yiran and Zheng, Wenyue and Cai, Tianle and Do, Xuan Long and Kawaguchi, Kenji and Goyal, Anirudh and Shieh, Michael Qizhe},
title = {Accelerating greedy coordinate gradient and general prompt optimization via probe sampling},
year = {2024},
isbn = {9798331314385},
publisher = {Curran Associates Inc.},
address = {Red Hook, NY, USA},
booktitle = {Proceedings of the 38th International Conference on Neural Information Processing Systems},
articleno = {1701},
numpages = {22},
location = {Vancouver, BC, Canada},
series = {NIPS '24}
}

@misc{metaphone_python,
  author       = {Andrew Collins and Contributors},
  title        = {Metaphone},
  howpublished = {\url{https://github.com/oubiwann/metaphone}},
  note         = {Python package on PyPI, BSD-licensed, Version 0.6, 2016-08-24}
}

@article{miller1995wordnet,
author = {Miller, George A.},
title = {WordNet: a lexical database for English},
year = {1995},
issue_date = {Nov. 1995},
publisher = {Association for Computing Machinery},
address = {New York, NY, USA},
volume = {38},
number = {11},
issn = {0001-0782},
url = {https://doi.org/10.1145/219717.219748},
doi = {10.1145/219717.219748},
journal = {Commun. ACM},
month = nov,
pages = {39–41},
numpages = {3}
}

@inproceedings{qi2021onionsimpleeffectivedefense,
    title = "{ONION}: A Simple and Effective Defense Against Textual Backdoor Attacks",
    author = "Qi, Fanchao  and
      Chen, Yangyi  and
      Li, Mukai  and
      Yao, Yuan  and
      Liu, Zhiyuan  and
      Sun, Maosong",
    editor = "Moens, Marie-Francine  and
      Huang, Xuanjing  and
      Specia, Lucia  and
      Yih, Scott Wen-tau",
    booktitle = "Proceedings of the 2021 Conference on Empirical Methods in Natural Language Processing",
    month = nov,
    year = "2021",
    address = "Online and Punta Cana, Dominican Republic",
    publisher = "Association for Computational Linguistics",
    url = "https://aclanthology.org/2021.emnlp-main.752/",
    doi = "10.18653/v1/2021.emnlp-main.752",
    pages = "9558--9566"
}

@misc{gao2020stripdefencetrojanattacks,
      title={STRIP: A Defence Against Trojan Attacks on Deep Neural Networks}, 
      author={Yansong Gao and Chang Xu and Derui Wang and Shiping Chen and Damith C. Ranasinghe and Surya Nepal},
      year={2020},
      eprint={1902.06531},
      archivePrefix={arXiv},
      primaryClass={cs.CR},
      url={https://arxiv.org/abs/1902.06531}, 
}

@inproceedings{he2023imbertmakingbertimmune,
    title = "{IMBERT}: Making {BERT} Immune to Insertion-based Backdoor Attacks",
    author = "He, Xuanli  and
      Wang, Jun  and
      Rubinstein, Benjamin  and
      Cohn, Trevor",
    editor = "Ovalle, Anaelia  and
      Chang, Kai-Wei  and
      Mehrabi, Ninareh  and
      Pruksachatkun, Yada  and
      Galystan, Aram  and
      Dhamala, Jwala  and
      Verma, Apurv  and
      Cao, Trista  and
      Kumar, Anoop  and
      Gupta, Rahul",
    booktitle = "Proceedings of the 3rd Workshop on Trustworthy Natural Language Processing (TrustNLP 2023)",
    month = jul,
    year = "2023",
    address = "Toronto, Canada",
    publisher = "Association for Computational Linguistics",
    url = "https://aclanthology.org/2023.trustnlp-1.25/",
    doi = "10.18653/v1/2023.trustnlp-1.25",
    pages = "287--301"
}

@inproceedings{huang2024compositebackdoorattackslarge,
    title = "Composite Backdoor Attacks Against Large Language Models",
    author = "Huang, Hai  and
      Zhao, Zhengyu  and
      Backes, Michael  and
      Shen, Yun  and
      Zhang, Yang",
    editor = "Duh, Kevin  and
      Gomez, Helena  and
      Bethard, Steven",
    booktitle = "Findings of the Association for Computational Linguistics: NAACL 2024",
    month = jun,
    year = "2024",
    address = "Mexico City, Mexico",
    publisher = "Association for Computational Linguistics",
    url = "https://aclanthology.org/2024.findings-naacl.94/",
    doi = "10.18653/v1/2024.findings-naacl.94",
    pages = "1459--1472"
}

@inproceedings{socher-etal-2013-recursive,
    title = "Recursive Deep Models for Semantic Compositionality Over a Sentiment Treebank",
    author = "Socher, Richard  and
      Perelygin, Alex  and
      Wu, Jean  and
      Chuang, Jason  and
      Manning, Christopher D.  and
      Ng, Andrew  and
      Potts, Christopher",
    editor = "Yarowsky, David  and
      Baldwin, Timothy  and
      Korhonen, Anna  and
      Livescu, Karen  and
      Bethard, Steven",
    booktitle = "Proceedings of the 2013 Conference on Empirical Methods in Natural Language Processing",
    month = oct,
    year = "2013",
    address = "Seattle, Washington, USA",
    publisher = "Association for Computational Linguistics",
    url = "https://aclanthology.org/D13-1170/",
    pages = "1631--1642"
}

@inproceedings{zampieri2019predictingtypetargetoffensive,
    title = "Predicting the Type and Target of Offensive Posts in Social Media",
    author = "Zampieri, Marcos  and
      Malmasi, Shervin  and
      Nakov, Preslav  and
      Rosenthal, Sara  and
      Farra, Noura  and
      Kumar, Ritesh",
    editor = "Burstein, Jill  and
      Doran, Christy  and
      Solorio, Thamar",
    booktitle = "Proceedings of the 2019 Conference of the North {A}merican Chapter of the Association for Computational Linguistics: Human Language Technologies, Volume 1 (Long and Short Papers)",
    month = jun,
    year = "2019",
    address = "Minneapolis, Minnesota",
    publisher = "Association for Computational Linguistics",
    url = "https://aclanthology.org/N19-1144/",
    doi = "10.18653/v1/N19-1144",
    pages = "1415--1420"
}

@inproceedings{Zhang2015CharacterlevelCN,
author = {Zhang, Xiang and Zhao, Junbo and LeCun, Yann},
title = {Character-level convolutional networks for text classification},
year = {2015},
publisher = {MIT Press},
address = {Cambridge, MA, USA},
booktitle = {Proceedings of the 29th International Conference on Neural Information Processing Systems - Volume 1},
pages = {649–657},
numpages = {9},
location = {Montreal, Canada},
series = {NIPS'15}
}

@inproceedings{devlin-etal-2019-bert,
    title = "{BERT}: Pre-training of Deep Bidirectional Transformers for Language Understanding",
    author = "Devlin, Jacob  and
      Chang, Ming-Wei  and
      Lee, Kenton  and
      Toutanova, Kristina",
    editor = "Burstein, Jill  and
      Doran, Christy  and
      Solorio, Thamar",
    booktitle = "Proceedings of the 2019 Conference of the North {A}merican Chapter of the Association for Computational Linguistics: Human Language Technologies, Volume 1 (Long and Short Papers)",
    month = jun,
    year = "2019",
    address = "Minneapolis, Minnesota",
    publisher = "Association for Computational Linguistics",
    url = "https://aclanthology.org/N19-1423/",
    doi = "10.18653/v1/N19-1423",
    pages = "4171--4186"
}

@misc{liu2019robertarobustlyoptimizedbert,
  title         = {RoBERTa: A Robustly Optimized BERT Pretraining Approach},
  author        = {Yinhan Liu and Myle Ott and Naman Goyal and Jingfei Du and Mandar Joshi and Danqi Chen and Omer Levy and Mike Lewis and Luke Zettlemoyer and Veselin Stoyanov},
  year          = {2019},
  eprint        = {1907.11692},
  archivePrefix = {arXiv},
  primaryClass  = {cs.CL},
  url           = {https://arxiv.org/abs/1907.11692}
}

@inproceedings {zhang2024instructionbackdoorattackscustomized,
author = {Rui Zhang and Hongwei Li and Rui Wen and Wenbo Jiang and Yuan Zhang and Michael Backes and Yun Shen and Yang Zhang},
title = {Instruction Backdoor Attacks Against Customized {LLMs}},
booktitle = {33rd USENIX Security Symposium (USENIX Security 24)},
year = {2024},
isbn = {978-1-939133-44-1},
address = {Philadelphia, PA},
pages = {1849--1866},
url = {https://www.usenix.org/conference/usenixsecurity24/presentation/zhang-rui},
publisher = {USENIX Association},
month = aug
}

@article{vanHeuven2014SUBTLEX,
  author   = {Walter J. B. van Heuven and Pawel Mandera and Emmanuel Keuleers and Marc Brysbaert},
  title    = {{SUBTLEX-UK}: A New and Improved Word Frequency Database for British English},
  journal  = {Quarterly Journal of Experimental Psychology},
  volume   = {67},
  number   = {6},
  pages    = {1176--1190},
  year     = {2014},
  doi      = {10.1080/17470218.2013.850521},
  note     = {PMID: 24417251},
  url      = {https://doi.org/10.1080/17470218.2013.850521}
}

@software{robyn_speer_2022_7199437,
  author       = {Robyn Speer},
  title        = {rspeer/wordfreq: v3.0},
  month        = sep,
  year         = 2022,
  publisher    = {Zenodo},
  version      = {v3.0.2},
  doi          = {10.5281/zenodo.7199437},
  url          = {https://doi.org/10.5281/zenodo.7199437}
}

@article{Hopkins_2025,
   title={AI Supply Chains: An Emerging Ecosystem of AI Actors, Products, and Services},
   volume={8},
   ISSN={3065-8365},
   url={http://dx.doi.org/10.1609/aies.v8i2.36628},
   DOI={10.1609/aies.v8i2.36628},
   number={2},
   journal={Proceedings of the AAAI/ACM Conference on AI, Ethics, and Society},
   publisher={Association for the Advancement of Artificial Intelligence (AAAI)},
   author={Hopkins, Aspen and Cen, Sarah H. and Struckman, Isabella and Ilyas, Andrew and Videgaray, Luis and Mądry, Aleksander},
   year={2025},
   month=Oct, pages={1266–1277} }

@techreport{PowellOswald2024,
  title        = {Assurance of Third-Party AI Systems for UK National Security: Research Report},
  author       = {Rosamund Powell and Marion Oswald},
  institution  = {The Alan Turing Institute},
  year         = {2024},
  address      = {London},
  url          = {https://cetas.turing.ac.uk/publications/assurance-third-party-ai-systems-uk-national-security},
  type         = {CETaS Research Report},
  month        = {January},
  note         = {Accessed 2025},
}

@techreport{Sherman2025SecuringAI,
  author       = {Justin Sherman},
  title        = {Securing Data in the AI Supply Chain},
  institution  = {Atlantic Council, Cyber Statecraft Initiative},
  year         = {2025},
  month        = {September},
  url          = {https://www.atlanticcouncil.org/in-depth-research-reports/issue-brief/securing-data-in-the-ai-supply-chain/},
  type         = {Issue Brief},
  note         = {Accessed 2026},
}

@inproceedings{zhang2025defensepromptinjectionattacks,
    title = "Defense against Prompt Injection Attacks via Mixture of Encodings",
    author = "Zhang, Ruiyi  and
      Sullivan, David  and
      Jackson, Kyle  and
      Xie, Pengtao  and
      Chen, Mei",
    editor = "Chiruzzo, Luis  and
      Ritter, Alan  and
      Wang, Lu",
    booktitle = "Proceedings of the 2025 Conference of the Nations of the Americas Chapter of the Association for Computational Linguistics: Human Language Technologies (Volume 2: Short Papers)",
    month = apr,
    year = "2025",
    address = "Albuquerque, New Mexico",
    publisher = "Association for Computational Linguistics",
    url = "https://aclanthology.org/2025.naacl-short.21/",
    doi = "10.18653/v1/2025.naacl-short.21",
    pages = "244--252",
    ISBN = "979-8-89176-190-2"
}

@misc{phi2_microsoft_2023,
  author = {Javaheripi, Mojan and Bubeck, Sébastien},
  title = {Phi-2: The surprising power of small language models},
  year = {2023},
  month = dec,
  howpublished = {https://www.microsoft.com/en-us/research/blog/phi-2-the-surprising-power-of-small-language-models/}
}

@article{deepseekai2025deepseekr1,
   title={DeepSeek-R1 incentivizes reasoning in LLMs through reinforcement learning},
   volume={645},
   ISSN={1476-4687},
   url={http://dx.doi.org/10.1038/s41586-025-09422-z},
   DOI={10.1038/s41586-025-09422-z},
   number={8081},
   journal={Nature},
   publisher={Springer Science and Business Media LLC},
   author={Guo, Daya and Yang, Dejian and Zhang, Haowei and Song, Junxiao and Wang, Peiyi and Zhu, Qihao and Xu, Runxin and Zhang, Ruoyu and Ma, Shirong and Bi, Xiao and Zhang, Xiaokang and Yu, Xingkai and Wu, Yu and Wu, Z. F. and Gou, Zhibin and Shao, Zhihong and Li, Zhuoshu and Gao, Ziyi and Liu, Aixin and Xue, Bing and Wang, Bingxuan and Wu, Bochao and Feng, Bei and Lu, Chengda and Zhao, Chenggang and Deng, Chengqi and Ruan, Chong and Dai, Damai and Chen, Deli and Ji, Dongjie and Li, Erhang and Lin, Fangyun and Dai, Fucong and Luo, Fuli and Hao, Guangbo and Chen, Guanting and Li, Guowei and Zhang, H. and Xu, Hanwei and Ding, Honghui and Gao, Huazuo and Qu, Hui and Li, Hui and Guo, Jianzhong and Li, Jiashi and Chen, Jingchang and Yuan, Jingyang and Tu, Jinhao and Qiu, Junjie and Li, Junlong and Cai, J. L. and Ni, Jiaqi and Liang, Jian and Chen, Jin and Dong, Kai and Hu, Kai and You, Kaichao and Gao, Kaige and Guan, Kang and Huang, Kexin and Yu, Kuai and Wang, Lean and Zhang, Lecong and Zhao, Liang and Wang, Litong and Zhang, Liyue and Xu, Lei and Xia, Leyi and Zhang, Mingchuan and Zhang, Minghua and Tang, Minghui and Zhou, Mingxu and Li, Meng and Wang, Miaojun and Li, Mingming and Tian, Ning and Huang, Panpan and Zhang, Peng and Wang, Qiancheng and Chen, Qinyu and Du, Qiushi and Ge, Ruiqi and Zhang, Ruisong and Pan, Ruizhe and Wang, Runji and Chen, R. J. and Jin, R. L. and Chen, Ruyi and Lu, Shanghao and Zhou, Shangyan and Chen, Shanhuang and Ye, Shengfeng and Wang, Shiyu and Yu, Shuiping and Zhou, Shunfeng and Pan, Shuting and Li, S. S. and Zhou, Shuang and Wu, Shaoqing and Yun, Tao and Pei, Tian and Sun, Tianyu and Wang, T. and Zeng, Wangding and Liu, Wen and Liang, Wenfeng and Gao, Wenjun and Yu, Wenqin and Zhang, Wentao and Xiao, W. L. and An, Wei and Liu, Xiaodong and Wang, Xiaohan and Chen, Xiaokang and Nie, Xiaotao and Cheng, Xin and Liu, Xin and Xie, Xin and Liu, Xingchao and Yang, Xinyu and Li, Xinyuan and Su, Xuecheng and Lin, Xuheng and Li, X. Q. and Jin, Xiangyue and Shen, Xiaojin and Chen, Xiaosha and Sun, Xiaowen and Wang, Xiaoxiang and Song, Xinnan and Zhou, Xinyi and Wang, Xianzu and Shan, Xinxia and Li, Y. K. and Wang, Y. Q. and Wei, Y. X. and Zhang, Yang and Xu, Yanhong and Li, Yao and Zhao, Yao and Sun, Yaofeng and Wang, Yaohui and Yu, Yi and Zhang, Yichao and Shi, Yifan and Xiong, Yiliang and He, Ying and Piao, Yishi and Wang, Yisong and Tan, Yixuan and Ma, Yiyang and Liu, Yiyuan and Guo, Yongqiang and Ou, Yuan and Wang, Yuduan and Gong, Yue and Zou, Yuheng and He, Yujia and Xiong, Yunfan and Luo, Yuxiang and You, Yuxiang and Liu, Yuxuan and Zhou, Yuyang and Zhu, Y. X. and Huang, Yanping and Li, Yaohui and Zheng, Yi and Zhu, Yuchen and Ma, Yunxian and Tang, Ying and Zha, Yukun and Yan, Yuting and Ren, Z. Z. and Ren, Zehui and Sha, Zhangli and Fu, Zhe and Xu, Zhean and Xie, Zhenda and Zhang, Zhengyan and Hao, Zhewen and Ma, Zhicheng and Yan, Zhigang and Wu, Zhiyu and Gu, Zihui and Zhu, Zijia and Liu, Zijun and Li, Zilin and Xie, Ziwei and Song, Ziyang and Pan, Zizheng and Huang, Zhen and Xu, Zhipeng and Zhang, Zhongyu and Zhang, Zhen},
   year={2025},
   month=Sept, pages={633–638} }

@misc{biderman2023pythiasuiteanalyzinglarge,
      title={Pythia: A Suite for Analyzing Large Language Models Across Training and Scaling}, 
      author={Stella Biderman and Hailey Schoelkopf and Quentin Anthony and Herbie Bradley and Kyle O'Brien and Eric Hallahan and Mohammad Aflah Khan and Shivanshu Purohit and USVSN Sai Prashanth and Edward Raff and Aviya Skowron and Lintang Sutawika and Oskar van der Wal},
      year={2023},
      eprint={2304.01373},
      archivePrefix={arXiv},
      primaryClass={cs.CL},
      url={https://arxiv.org/abs/2304.01373}, 
}

@misc{touvron2023llama2openfoundation,
      title={Llama 2: Open Foundation and Fine-Tuned Chat Models}, 
      author={Hugo Touvron and Louis Martin and Kevin Stone and Peter Albert and Amjad Almahairi and Yasmine Babaei and Nikolay Bashlykov and Soumya Batra and Prajjwal Bhargava and Shruti Bhosale and Dan Bikel and Lukas Blecher and Cristian Canton Ferrer and Moya Chen and Guillem Cucurull and David Esiobu and Jude Fernandes and Jeremy Fu and Wenyin Fu and Brian Fuller and Cynthia Gao and Vedanuj Goswami and Naman Goyal and Anthony Hartshorn and Saghar Hosseini and Rui Hou and Hakan Inan and Marcin Kardas and Viktor Kerkez and Madian Khabsa and Isabel Kloumann and Artem Korenev and Punit Singh Koura and Marie-Anne Lachaux and Thibaut Lavril and Jenya Lee and Diana Liskovich and Yinghai Lu and Yuning Mao and Xavier Martinet and Todor Mihaylov and Pushkar Mishra and Igor Molybog and Yixin Nie and Andrew Poulton and Jeremy Reizenstein and Rashi Rungta and Kalyan Saladi and Alan Schelten and Ruan Silva and Eric Michael Smith and Ranjan Subramanian and Xiaoqing Ellen Tan and Binh Tang and Ross Taylor and Adina Williams and Jian Xiang Kuan and Puxin Xu and Zheng Yan and Iliyan Zarov and Yuchen Zhang and Angela Fan and Melanie Kambadur and Sharan Narang and Aurelien Rodriguez and Robert Stojnic and Sergey Edunov and Thomas Scialom},
      year={2023},
      eprint={2307.09288},
      archivePrefix={arXiv},
      primaryClass={cs.CL},
      url={https://arxiv.org/abs/2307.09288}, 
}

@misc{jiang2023mistral7b,
      title={Mistral 7B}, 
      author={Albert Q. Jiang and Alexandre Sablayrolles and Arthur Mensch and Chris Bamford and Devendra Singh Chaplot and Diego de las Casas and Florian Bressand and Gianna Lengyel and Guillaume Lample and Lucile Saulnier and Lélio Renard Lavaud and Marie-Anne Lachaux and Pierre Stock and Teven Le Scao and Thibaut Lavril and Thomas Wang and Timothée Lacroix and William El Sayed},
      year={2023},
      eprint={2310.06825},
      archivePrefix={arXiv},
      primaryClass={cs.CL},
      url={https://arxiv.org/abs/2310.06825}, 
}

@inproceedings{yang-etal-2021-rethinking,
    title = "Rethinking Stealthiness of Backdoor Attack against {NLP} Models",
    author = "Yang, Wenkai  and
      Lin, Yankai  and
      Li, Peng  and
      Zhou, Jie  and
      Sun, Xu",
    editor = "Zong, Chengqing  and
      Xia, Fei  and
      Li, Wenjie  and
      Navigli, Roberto",
    booktitle = "Proceedings of the 59th Annual Meeting of the Association for Computational Linguistics and the 11th International Joint Conference on Natural Language Processing (Volume 1: Long Papers)",
    month = aug,
    year = "2021",
    address = "Online",
    publisher = "Association for Computational Linguistics",
    url = "https://aclanthology.org/2021.acl-long.431/",
    doi = "10.18653/v1/2021.acl-long.431",
    pages = "5543--5557"
}

@misc{gemmateam2024gemma2improvingopen,
      title={Gemma 2: Improving Open Language Models at a Practical Size}, 
      author={Gemma Team and Morgane Riviere and Shreya Pathak and Pier Giuseppe Sessa and Cassidy Hardin and Surya Bhupatiraju and Léonard Hussenot and Thomas Mesnard and Bobak Shahriari and Alexandre Ramé and Johan Ferret and Peter Liu and Pouya Tafti and Abe Friesen and Michelle Casbon and Sabela Ramos and Ravin Kumar and Charline Le Lan and Sammy Jerome and Anton Tsitsulin and Nino Vieillard and Piotr Stanczyk and Sertan Girgin and Nikola Momchev and Matt Hoffman and Shantanu Thakoor and Jean-Bastien Grill and Behnam Neyshabur and Olivier Bachem and Alanna Walton and Aliaksei Severyn and Alicia Parrish and Aliya Ahmad and Allen Hutchison and Alvin Abdagic and Amanda Carl and Amy Shen and Andy Brock and Andy Coenen and Anthony Laforge and Antonia Paterson and Ben Bastian and Bilal Piot and Bo Wu and Brandon Royal and Charlie Chen and Chintu Kumar and Chris Perry and Chris Welty and Christopher A. Choquette-Choo and Danila Sinopalnikov and David Weinberger and Dimple Vijaykumar and Dominika Rogozińska and Dustin Herbison and Elisa Bandy and Emma Wang and Eric Noland and Erica Moreira and Evan Senter and Evgenii Eltyshev and Francesco Visin and Gabriel Rasskin and Gary Wei and Glenn Cameron and Gus Martins and Hadi Hashemi and Hanna Klimczak-Plucińska and Harleen Batra and Harsh Dhand and Ivan Nardini and Jacinda Mein and Jack Zhou and James Svensson and Jeff Stanway and Jetha Chan and Jin Peng Zhou and Joana Carrasqueira and Joana Iljazi and Jocelyn Becker and Joe Fernandez and Joost van Amersfoort and Josh Gordon and Josh Lipschultz and Josh Newlan and Ju-yeong Ji and Kareem Mohamed and Kartikeya Badola and Kat Black and Katie Millican and Keelin McDonell and Kelvin Nguyen and Kiranbir Sodhia and Kish Greene and Lars Lowe Sjoesund and Lauren Usui and Laurent Sifre and Lena Heuermann and Leticia Lago and Lilly McNealus and Livio Baldini Soares and Logan Kilpatrick and Lucas Dixon and Luciano Martins and Machel Reid and Manvinder Singh and Mark Iverson and Martin Görner and Mat Velloso and Mateo Wirth and Matt Davidow and Matt Miller and Matthew Rahtz and Matthew Watson and Meg Risdal and Mehran Kazemi and Michael Moynihan and Ming Zhang and Minsuk Kahng and Minwoo Park and Mofi Rahman and Mohit Khatwani and Natalie Dao and Nenshad Bardoliwalla and Nesh Devanathan and Neta Dumai and Nilay Chauhan and Oscar Wahltinez and Pankil Botarda and Parker Barnes and Paul Barham and Paul Michel and Pengchong Jin and Petko Georgiev and Phil Culliton and Pradeep Kuppala and Ramona Comanescu and Ramona Merhej and Reena Jana and Reza Ardeshir Rokni and Rishabh Agarwal and Ryan Mullins and Samaneh Saadat and Sara Mc Carthy and Sarah Cogan and Sarah Perrin and Sébastien M. R. Arnold and Sebastian Krause and Shengyang Dai and Shruti Garg and Shruti Sheth and Sue Ronstrom and Susan Chan and Timothy Jordan and Ting Yu and Tom Eccles and Tom Hennigan and Tomas Kocisky and Tulsee Doshi and Vihan Jain and Vikas Yadav and Vilobh Meshram and Vishal Dharmadhikari and Warren Barkley and Wei Wei and Wenming Ye and Woohyun Han and Woosuk Kwon and Xiang Xu and Zhe Shen and Zhitao Gong and Zichuan Wei and Victor Cotruta and Phoebe Kirk and Anand Rao and Minh Giang and Ludovic Peran and Tris Warkentin and Eli Collins and Joelle Barral and Zoubin Ghahramani and Raia Hadsell and D. Sculley and Jeanine Banks and Anca Dragan and Slav Petrov and Oriol Vinyals and Jeff Dean and Demis Hassabis and Koray Kavukcuoglu and Clement Farabet and Elena Buchatskaya and Sebastian Borgeaud and Noah Fiedel and Armand Joulin and Kathleen Kenealy and Robert Dadashi and Alek Andreev},
      year={2024},
      eprint={2408.00118},
      archivePrefix={arXiv},
      primaryClass={cs.CL},
      url={https://arxiv.org/abs/2408.00118}, 
}

@misc{abdin2024phi4technicalreport,
      title={Phi-4 Technical Report}, 
      author={Marah Abdin and Jyoti Aneja and Harkirat Behl and Sébastien Bubeck and Ronen Eldan and Suriya Gunasekar and Michael Harrison and Russell J. Hewett and Mojan Javaheripi and Piero Kauffmann and James R. Lee and Yin Tat Lee and Yuanzhi Li and Weishung Liu and Caio C. T. Mendes and Anh Nguyen and Eric Price and Gustavo de Rosa and Olli Saarikivi and Adil Salim and Shital Shah and Xin Wang and Rachel Ward and Yue Wu and Dingli Yu and Cyril Zhang and Yi Zhang},
      year={2024},
      eprint={2412.08905},
      archivePrefix={arXiv},
      primaryClass={cs.CL},
      url={https://arxiv.org/abs/2412.08905}, 
}

@misc{bai2023qwentechnicalreport,
      title={Qwen Technical Report}, 
      author={Jinze Bai and Shuai Bai and Yunfei Chu and Zeyu Cui and Kai Dang and Xiaodong Deng and Yang Fan and Wenbin Ge and Yu Han and Fei Huang and Binyuan Hui and Luo Ji and Mei Li and Junyang Lin and Runji Lin and Dayiheng Liu and Gao Liu and Chengqiang Lu and Keming Lu and Jianxin Ma and Rui Men and Xingzhang Ren and Xuancheng Ren and Chuanqi Tan and Sinan Tan and Jianhong Tu and Peng Wang and Shijie Wang and Wei Wang and Shengguang Wu and Benfeng Xu and Jin Xu and An Yang and Hao Yang and Jian Yang and Shusheng Yang and Yang Yao and Bowen Yu and Hongyi Yuan and Zheng Yuan and Jianwei Zhang and Xingxuan Zhang and Yichang Zhang and Zhenru Zhang and Chang Zhou and Jingren Zhou and Xiaohuan Zhou and Tianhang Zhu},
      year={2023},
      eprint={2309.16609},
      archivePrefix={arXiv},
      primaryClass={cs.CL},
      url={https://arxiv.org/abs/2309.16609}, 
}

@misc{yang2025qwen3technicalreport,
      title={Qwen3 Technical Report}, 
      author={An Yang and Anfeng Li and Baosong Yang and Beichen Zhang and Binyuan Hui and Bo Zheng and Bowen Yu and Chang Gao and Chengen Huang and Chenxu Lv and Chujie Zheng and Dayiheng Liu and Fan Zhou and Fei Huang and Feng Hu and Hao Ge and Haoran Wei and Huan Lin and Jialong Tang and Jian Yang and Jianhong Tu and Jianwei Zhang and Jianxin Yang and Jiaxi Yang and Jing Zhou and Jingren Zhou and Junyang Lin and Kai Dang and Keqin Bao and Kexin Yang and Le Yu and Lianghao Deng and Mei Li and Mingfeng Xue and Mingze Li and Pei Zhang and Peng Wang and Qin Zhu and Rui Men and Ruize Gao and Shixuan Liu and Shuang Luo and Tianhao Li and Tianyi Tang and Wenbiao Yin and Xingzhang Ren and Xinyu Wang and Xinyu Zhang and Xuancheng Ren and Yang Fan and Yang Su and Yichang Zhang and Yinger Zhang and Yu Wan and Yuqiong Liu and Zekun Wang and Zeyu Cui and Zhenru Zhang and Zhipeng Zhou and Zihan Qiu},
      year={2025},
      eprint={2505.09388},
      archivePrefix={arXiv},
      primaryClass={cs.CL},
      url={https://arxiv.org/abs/2505.09388}, 
}

@misc{qwen2025qwen25technicalreport,
      title={Qwen2.5 Technical Report}, 
      author={An Yang and Baosong Yang and Beichen Zhang and Binyuan Hui and Bo Zheng and Bowen Yu and Chengyuan Li and Dayiheng Liu and Fei Huang and Haoran Wei and Huan Lin and Jian Yang and Jianhong Tu and Jianwei Zhang and Jianxin Yang and Jiaxi Yang and Jingren Zhou and Junyang Lin and Kai Dang and Keming Lu and Keqin Bao and Kexin Yang and Le Yu and Mei Li and Mingfeng Xue and Pei Zhang and Qin Zhu and Rui Men and Runji Lin and Tianhao Li and Tianyi Tang and Tingyu Xia and Xingzhang Ren and Xuancheng Ren and Yang Fan and Yang Su and Yichang Zhang and Yu Wan and Yuqiong Liu and Zeyu Cui and Zhenru Zhang and Zihan Qiu},
      year={2025},
      eprint={2412.15115},
      archivePrefix={arXiv},
      primaryClass={cs.CL},
      url={https://arxiv.org/abs/2412.15115}, 
}

@misc{grattafiori2024llama3herdmodels,
      title={The Llama 3 Herd of Models}, 
      author={Aaron Grattafiori and Abhimanyu Dubey and Abhinav Jauhri and Abhinav Pandey and Abhishek Kadian and Ahmad Al-Dahle and Aiesha Letman and Akhil Mathur and Alan Schelten and Alex Vaughan and Amy Yang and Angela Fan and Anirudh Goyal and Anthony Hartshorn and Aobo Yang and Archi Mitra and Archie Sravankumar and Artem Korenev and Arthur Hinsvark and Arun Rao and Aston Zhang and Aurelien Rodriguez and Austen Gregerson and Ava Spataru and Baptiste Roziere and Bethany Biron and Binh Tang and Bobbie Chern and Charlotte Caucheteux and Chaya Nayak and Chloe Bi and Chris Marra and Chris McConnell and Christian Keller and Christophe Touret and Chunyang Wu and Corinne Wong and Cristian Canton Ferrer and Cyrus Nikolaidis and Damien Allonsius and Daniel Song and Danielle Pintz and Danny Livshits and Danny Wyatt and David Esiobu and Dhruv Choudhary and Dhruv Mahajan and Diego Garcia-Olano and Diego Perino and Dieuwke Hupkes and Egor Lakomkin and Ehab AlBadawy and Elina Lobanova and Emily Dinan and Eric Michael Smith and Filip Radenovic and Francisco Guzmán and Frank Zhang and Gabriel Synnaeve and Gabrielle Lee and Georgia Lewis Anderson and Govind Thattai and Graeme Nail and Gregoire Mialon and Guan Pang and Guillem Cucurell and Hailey Nguyen and Hannah Korevaar and Hu Xu and Hugo Touvron and Iliyan Zarov and Imanol Arrieta Ibarra and Isabel Kloumann and Ishan Misra and Ivan Evtimov and Jack Zhang and Jade Copet and Jaewon Lee and Jan Geffert and Jana Vranes and Jason Park and Jay Mahadeokar and Jeet Shah and Jelmer van der Linde and Jennifer Billock and Jenny Hong and Jenya Lee and Jeremy Fu and Jianfeng Chi and Jianyu Huang and Jiawen Liu and Jie Wang and Jiecao Yu and Joanna Bitton and Joe Spisak and Jongsoo Park and Joseph Rocca and Joshua Johnstun and Joshua Saxe and Junteng Jia and Kalyan Vasuden Alwala and Karthik Prasad and Kartikeya Upasani and Kate Plawiak and Ke Li and Kenneth Heafield and Kevin Stone and Khalid El-Arini and Krithika Iyer and Kshitiz Malik and Kuenley Chiu and Kunal Bhalla and Kushal Lakhotia and Lauren Rantala-Yeary and Laurens van der Maaten and Lawrence Chen and Liang Tan and Liz Jenkins and Louis Martin and Lovish Madaan and Lubo Malo and Lukas Blecher and Lukas Landzaat and Luke de Oliveira and Madeline Muzzi and Mahesh Pasupuleti and Mannat Singh and Manohar Paluri and Marcin Kardas and Maria Tsimpoukelli and Mathew Oldham and Mathieu Rita and Maya Pavlova and Melanie Kambadur and Mike Lewis and Min Si and Mitesh Kumar Singh and Mona Hassan and Naman Goyal and Narjes Torabi and Nikolay Bashlykov and Nikolay Bogoychev and Niladri Chatterji and Ning Zhang and Olivier Duchenne and Onur Çelebi and Patrick Alrassy and Pengchuan Zhang and Pengwei Li and Petar Vasic and Peter Weng and Prajjwal Bhargava and Pratik Dubal and Praveen Krishnan and Punit Singh Koura and Puxin Xu and Qing He and Qingxiao Dong and Ragavan Srinivasan and Raj Ganapathy and Ramon Calderer and Ricardo Silveira Cabral and Robert Stojnic and Roberta Raileanu and Rohan Maheswari and Rohit Girdhar and Rohit Patel and Romain Sauvestre and Ronnie Polidoro and Roshan Sumbaly and Ross Taylor and Ruan Silva and Rui Hou and Rui Wang and Saghar Hosseini and Sahana Chennabasappa and Sanjay Singh and Sean Bell and Seohyun Sonia Kim and Sergey Edunov and Shaoliang Nie and Sharan Narang and Sharath Raparthy and Sheng Shen and Shengye Wan and Shruti Bhosale and Shun Zhang and Simon Vandenhende and Soumya Batra and Spencer Whitman and Sten Sootla and Stephane Collot and Suchin Gururangan and Sydney Borodinsky and Tamar Herman and Tara Fowler and Tarek Sheasha and Thomas Georgiou and Thomas Scialom and Tobias Speckbacher and Todor Mihaylov and Tong Xiao and Ujjwal Karn and Vedanuj Goswami and Vibhor Gupta and Vignesh Ramanathan and Viktor Kerkez and Vincent Gonguet and Virginie Do and Vish Vogeti and Vítor Albiero and Vladan Petrovic and Weiwei Chu and Wenhan Xiong and Wenyin Fu and Whitney Meers and Xavier Martinet and Xiaodong Wang and Xiaofang Wang and Xiaoqing Ellen Tan and Xide Xia and Xinfeng Xie and Xuchao Jia and Xuewei Wang and Yaelle Goldschlag and Yashesh Gaur and Yasmine Babaei and Yi Wen and Yiwen Song and Yuchen Zhang and Yue Li and Yuning Mao and Zacharie Delpierre Coudert and Zheng Yan and Zhengxing Chen and Zoe Papakipos and Aaditya Singh and Aayushi Srivastava and Abha Jain and Adam Kelsey and Adam Shajnfeld and Adithya Gangidi and Adolfo Victoria and Ahuva Goldstand and Ajay Menon and Ajay Sharma and Alex Boesenberg and Alexei Baevski and Allie Feinstein and Amanda Kallet and Amit Sangani and Amos Teo and Anam Yunus and Andrei Lupu and Andres Alvarado and Andrew Caples and Andrew Gu and Andrew Ho and Andrew Poulton and Andrew Ryan and Ankit Ramchandani and Annie Dong and Annie Franco and Anuj Goyal and Aparajita Saraf and Arkabandhu Chowdhury and Ashley Gabriel and Ashwin Bharambe and Assaf Eisenman and Azadeh Yazdan and Beau James and Ben Maurer and Benjamin Leonhardi and Bernie Huang and Beth Loyd and Beto De Paola and Bhargavi Paranjape and Bing Liu and Bo Wu and Boyu Ni and Braden Hancock and Bram Wasti and Brandon Spence and Brani Stojkovic and Brian Gamido and Britt Montalvo and Carl Parker and Carly Burton and Catalina Mejia and Ce Liu and Changhan Wang and Changkyu Kim and Chao Zhou and Chester Hu and Ching-Hsiang Chu and Chris Cai and Chris Tindal and Christoph Feichtenhofer and Cynthia Gao and Damon Civin and Dana Beaty and Daniel Kreymer and Daniel Li and David Adkins and David Xu and Davide Testuggine and Delia David and Devi Parikh and Diana Liskovich and Didem Foss and Dingkang Wang and Duc Le and Dustin Holland and Edward Dowling and Eissa Jamil and Elaine Montgomery and Eleonora Presani and Emily Hahn and Emily Wood and Eric-Tuan Le and Erik Brinkman and Esteban Arcaute and Evan Dunbar and Evan Smothers and Fei Sun and Felix Kreuk and Feng Tian and Filippos Kokkinos and Firat Ozgenel and Francesco Caggioni and Frank Kanayet and Frank Seide and Gabriela Medina Florez and Gabriella Schwarz and Gada Badeer and Georgia Swee and Gil Halpern and Grant Herman and Grigory Sizov and Guangyi and Zhang and Guna Lakshminarayanan and Hakan Inan and Hamid Shojanazeri and Han Zou and Hannah Wang and Hanwen Zha and Haroun Habeeb and Harrison Rudolph and Helen Suk and Henry Aspegren and Hunter Goldman and Hongyuan Zhan and Ibrahim Damlaj and Igor Molybog and Igor Tufanov and Ilias Leontiadis and Irina-Elena Veliche and Itai Gat and Jake Weissman and James Geboski and James Kohli and Janice Lam and Japhet Asher and Jean-Baptiste Gaya and Jeff Marcus and Jeff Tang and Jennifer Chan and Jenny Zhen and Jeremy Reizenstein and Jeremy Teboul and Jessica Zhong and Jian Jin and Jingyi Yang and Joe Cummings and Jon Carvill and Jon Shepard and Jonathan McPhie and Jonathan Torres and Josh Ginsburg and Junjie Wang and Kai Wu and Kam Hou U and Karan Saxena and Kartikay Khandelwal and Katayoun Zand and Kathy Matosich and Kaushik Veeraraghavan and Kelly Michelena and Keqian Li and Kiran Jagadeesh and Kun Huang and Kunal Chawla and Kyle Huang and Lailin Chen and Lakshya Garg and Lavender A and Leandro Silva and Lee Bell and Lei Zhang and Liangpeng Guo and Licheng Yu and Liron Moshkovich and Luca Wehrstedt and Madian Khabsa and Manav Avalani and Manish Bhatt and Martynas Mankus and Matan Hasson and Matthew Lennie and Matthias Reso and Maxim Groshev and Maxim Naumov and Maya Lathi and Meghan Keneally and Miao Liu and Michael L. Seltzer and Michal Valko and Michelle Restrepo and Mihir Patel and Mik Vyatskov and Mikayel Samvelyan and Mike Clark and Mike Macey and Mike Wang and Miquel Jubert Hermoso and Mo Metanat and Mohammad Rastegari and Munish Bansal and Nandhini Santhanam and Natascha Parks and Natasha White and Navyata Bawa and Nayan Singhal and Nick Egebo and Nicolas Usunier and Nikhil Mehta and Nikolay Pavlovich Laptev and Ning Dong and Norman Cheng and Oleg Chernoguz and Olivia Hart and Omkar Salpekar and Ozlem Kalinli and Parkin Kent and Parth Parekh and Paul Saab and Pavan Balaji and Pedro Rittner and Philip Bontrager and Pierre Roux and Piotr Dollar and Polina Zvyagina and Prashant Ratanchandani and Pritish Yuvraj and Qian Liang and Rachad Alao and Rachel Rodriguez and Rafi Ayub and Raghotham Murthy and Raghu Nayani and Rahul Mitra and Rangaprabhu Parthasarathy and Raymond Li and Rebekkah Hogan and Robin Battey and Rocky Wang and Russ Howes and Ruty Rinott and Sachin Mehta and Sachin Siby and Sai Jayesh Bondu and Samyak Datta and Sara Chugh and Sara Hunt and Sargun Dhillon and Sasha Sidorov and Satadru Pan and Saurabh Mahajan and Saurabh Verma and Seiji Yamamoto and Sharadh Ramaswamy and Shaun Lindsay and Shaun Lindsay and Sheng Feng and Shenghao Lin and Shengxin Cindy Zha and Shishir Patil and Shiva Shankar and Shuqiang Zhang and Shuqiang Zhang and Sinong Wang and Sneha Agarwal and Soji Sajuyigbe and Soumith Chintala and Stephanie Max and Stephen Chen and Steve Kehoe and Steve Satterfield and Sudarshan Govindaprasad and Sumit Gupta and Summer Deng and Sungmin Cho and Sunny Virk and Suraj Subramanian and Sy Choudhury and Sydney Goldman and Tal Remez and Tamar Glaser and Tamara Best and Thilo Koehler and Thomas Robinson and Tianhe Li and Tianjun Zhang and Tim Matthews and Timothy Chou and Tzook Shaked and Varun Vontimitta and Victoria Ajayi and Victoria Montanez and Vijai Mohan and Vinay Satish Kumar and Vishal Mangla and Vlad Ionescu and Vlad Poenaru and Vlad Tiberiu Mihailescu and Vladimir Ivanov and Wei Li and Wenchen Wang and Wenwen Jiang and Wes Bouaziz and Will Constable and Xiaocheng Tang and Xiaojian Wu and Xiaolan Wang and Xilun Wu and Xinbo Gao and Yaniv Kleinman and Yanjun Chen and Ye Hu and Ye Jia and Ye Qi and Yenda Li and Yilin Zhang and Ying Zhang and Yossi Adi and Youngjin Nam and Yu and Wang and Yu Zhao and Yuchen Hao and Yundi Qian and Yunlu Li and Yuzi He and Zach Rait and Zachary DeVito and Zef Rosnbrick and Zhaoduo Wen and Zhenyu Yang and Zhiwei Zhao and Zhiyu Ma},
      year={2024},
      eprint={2407.21783},
      archivePrefix={arXiv},
      primaryClass={cs.AI},
      url={https://arxiv.org/abs/2407.21783}, 
}

\clearpage

\appendix

\section{Exploratory Human Audit}
\label{app:human_audit}

We conducted a small expert audit using the RoBERTa-base / SST-2 / `James Bond'' setting from Table~\ref{tab:final_correct}. We inserted a single \textsc{SteganoPoison} that had successfully survived the complete automated defense pipeline (`The micromanagement from Alberta creates endless conflict.'') into a set of 99 clean SST-2 examples. Six machine-learning PhD students, all of whom had carefully read the paper and were familiar with SST-2, were asked to identify any samples they believed to be poisoned. Participants were informed that SST-2 naturally contains noisy, awkward, and occasionally semantically unusual examples, but were not told whether any samples had been modified, how many poisoned samples (if any) were present, or which examples might have been altered. They were given unlimited time and did not have access to the labels. Only 2/6 participants identified the poisoned sample.

We do not draw conclusions from this pilot because its small scale and artificially high poison prevalence are insufficient to support statistically meaningful claims.

\section{Stealth and Anti-Forensic Properties}
\label{sec:stealth_properties}

To remain effective under pre-training data-poisoning defenses, which may be heterogeneous and defender-favoring, the adversary must design poisons that simultaneously satisfy constraints on visibility, optimization efficiency, and forensic trace reduction. Our framework achieves this through two distinct mechanisms that operate at different layers of the pipeline.

First, at the text-content level, \textsc{SteganoPoisons} contain no explicit inference-time trigger, maintain linguistic fluency, and encode the training-time payload across ordinary tokens rather than concentrating it in a small set of trigger-aligned features. Because no individual token carries a concentrated signal, defenses that rely on localized statistical, fluency, perturbation, or gradient-based irregularities have limited evidence with which to identify the malicious training examples.

Second, at the architectural level, because poison optimization operates within a specific tokenizer and its corresponding input embedding space, the resulting \textsc{SteganoPoisons} are effectively tokenizer-locked. As demonstrated in our cross-tokenizer evaluation, \textsc{SteganoPoisons} transfer across model weights and scales when the tokenizer is shared, but do not transfer when used to train models with different tokenizers. This dependency creates an additional anti-forensic barrier: an auditing pipeline using a surrogate with a mismatched tokenizer cannot reliably recover the backdoor signal encoded during poison construction.

Together, these properties make attribution to individual training examples substantially more difficult. A data auditor cannot rely on either a concentrated trigger-associated signature or cross-tokenizer transfer, while the \textsc{SteganoPoisons} themselves remain fluent and constitute only a small fraction of the training corpus.

\section{Defense Configurations}
\label{sec:defense-settings}

\textbf{ONION}~\cite{qi2021onionsimpleeffectivedefense} is instantiated in a more aggressive form as a perplexity-based, sample-level detector that probes sentence fluency under token deletion. Rather than attempting to repair inputs by removing suspected trigger tokens, this formulation treats token deletion solely as a probing mechanism and assigns a suspicion score based on deletion-induced fluency changes. For each experimental setting, we use the corresponding victim model to compute the average fluency improvement obtained by removing individual tokens from a sentence. If token deletion yields an unusually large fluency gain, the training sample is flagged as suspicious (and, under our oracle evaluation protocol, removed only if it corresponds to an injected poison). Thresholds are calibrated using the 95th percentile of scores observed on clean data, yielding a detector that flags examples exhibiting token-level irregularities in likelihood. In our setting, \textsc{SteganoPoisons} are explicitly constructed to evade this signal: poison optimization constrains sentence-level perplexity relative to the clean-data distribution under the diagnostic model, while all tokens in seed poisons and all replacement tokens are restricted to valid dictionary words, eliminating rare, malformed, or synthetic tokens.

Because \textsc{SteganoPoisons} contain no localized likelihood spikes and no deletion-sensitive fluency artifacts, removing any single token produces negligible or inconsistent changes in perplexity, causing the ONION deletion-based suspicion score to remain below the clean-calibrated threshold despite aggressive probing.

\textbf{STRIP}~\cite{gao2020stripdefencetrojanattacks} is instantiated as an offline, pre-training \emph{sample-level detector} rather than an inference-time defense.
Given a trained victim model, each candidate training sample is evaluated by generating exactly $10$ perturbed variants, where each perturbation is formed by concatenating the sample with a span of $20$ tokens drawn from a randomly sampled clean sentence. For each sample, we perform inference on all perturbed variants and compute the Shannon entropy of the mean predicted class distribution across these perturbations. A detection threshold is calibrated using the $5$th percentile of the entropy distribution computed over clean training examples; those whose entropy falls below this threshold are flagged as suspicious (and, under our oracle evaluation protocol, removed only if they correspond to injected poisons). Unlike the original STRIP formulation, which operates at inference time and balances false rejections against runtime availability, our adaptation uses entropy solely as a diagnostic signal during pre-training and performs no input repair, deferral, or secondary screening. Under this pre-training formulation, STRIP is ineffective against \textsc{SteganoBackdoor}. When the victim model is used as the STRIP surrogate, \textsc{SteganoPoisons} do not activate backdoor behavior during screening because they contain no inference-time trigger tokens. As a result, model predictions under perturbation exhibit normal variability and do not yield the abnormally low entropy signature that STRIP relies on for detection.

\textbf{IMBERT}~\cite{he2023imbertmakingbertimmune} is adapted from a token-level inference-time defense into a gradient-based, sample-level diagnostic for pre-training data curation. Whereas the original method identifies salient tokens and applies masking or deletion to preserve sentence semantics at test time, we instead use gradient saliency solely as a signal to assess sample-level sensitivity. For each training instance, we compute gradients of the pseudo-label cross-entropy loss with respect to input token embeddings using the victim model, take token-wise $\ell_2$ norms over the embedding dimension, and aggregate them by averaging across all tokens to obtain a single scalar score. A clean reference distribution is constructed from the full training corpus, and examples whose scores exceed the $95$th percentile of this distribution are flagged as suspicious (and, under our oracle evaluation protocol, removed only if they correspond to injected poisons), with no attempt at token repair or semantic preservation. This adaptation shifts IMBERT from accuracy-preserving inference-time mitigation to diagnostic poisoned-sample detection in a pre-training setting, but remains ineffective against \textsc{SteganoBackdoor} because the backdoor payload is distributed across tokens rather than concentrated in a small set of trigger-aligned features, preventing any individual token from exhibiting abnormally large gradient norms and causing sample-level saliency scores to remain within the clean-calibrated threshold despite aggressive probing.

\textbf{SCPD}~\cite{HIDDENKILLER} is instantiated as a sample-level syntactic consistency diagnostic for pre-training data curation.
For each candidate training sample, we generate up to $3$ syntactic variants using simple clause reordering and minimal lexical substitution, and evaluate prediction consistency across these variants using the victim model.
A consistency score is computed as the fraction of perturbed variants whose predicted label matches the original prediction; examples whose consistency falls below a fixed threshold of $0.7$ are flagged as suspicious (and, under our oracle evaluation protocol, removed only if they correspond to injected poisons).
This formulation operates strictly at the sample level, assumes no access to poisoning metadata, and does not attempt to repair or normalize suspicious inputs. SCPD is ineffective against \textsc{SteganoBackdoor} because \textsc{SteganoPoisons} do not rely on inference-time syntactic triggers, and syntactic perturbations therefore do not induce the prediction instability this defense targets.

\textbf{maxEntropy}~\cite{cheng-etal-2025-synghost} is instantiated as an entropy-based, sample-level diagnostic for pre-training data curation. Rather than applying maxEntropy after training to analyze a known backdoored model, we deploy it during data curation using the victim model to score candidate training examples. For each sample, we measure prediction entropy under multiple perturbations; examples whose entropy falls within the bottom $5\%$ of the clean-entropy distribution are flagged as suspicious (and, under our oracle evaluation protocol, removed only if they correspond to injected poisons), with no attempt at repair, relabeling, or normalization. This formulation treats unusually stable predictions under perturbation as a sufficient signal for poison detection. maxEntropy is ineffective against \textsc{SteganoBackdoor} because \textsc{SteganoPoisons} do not contain inference-time triggers and therefore do not activate backdoor behavior during probing, causing their entropy scores to remain within the clean-calibrated range.

\section{Hyperparameters}
\label{app:hyper}
All experiments in this work use the default hyperparameter values listed in Table~\ref{tab:method-hparams}.

Because absolute perplexity values and embedding similarity statistics vary substantially across model architectures and tokenizers, both the fluency and overlap terms are calibrated separately for each diagnostic model using empirical distributions computed on the clean training data.
Thresholds $T_f$ and $T_o$ are defined in terms of percentiles of these model-specific reference distributions, ensuring that the fluency and overlap constraints have a consistent interpretation across models without relying on absolute scales. This calibration affects only the relative scaling of the auxiliary penalties and does not change the qualitative behavior of the optimization objective.

\begin{table}[t]
\centering
\scriptsize
\setlength{\tabcolsep}{6pt}
\renewcommand{\arraystretch}{1.15}
\resizebox{\columnwidth}{!}{
\begin{tabular}{p{0.36\columnwidth} p{0.24\columnwidth} p{0.32\columnwidth}}
\hline
\textbf{Component} & \textbf{Param.} & \textbf{Value} \\
\hline
\textit{SteganoPoison Objective} & & \\
Payload weight & $\lambda_p$ & $1.0$ \\
Fluency weight & $\lambda_f$ & $0.01$ \\
Overlap weight & $\lambda_o$ & $0.01$ \\
\hline
\textit{Payload Term $\mathcal{L}_{\mathrm{p}}$} & & \\
Inner-step LR & $\eta$ & $2 \times 10^{-2}$ \\
Probe set size & $|\mathcal{T}|$ & $128$ \\
\hline
\textit{Fluency Term $\mathcal{L}_{\mathrm{f}}$} & & \\
Fluency threshold & $T_f$ & clean PPL (10th pct.) \\
Sharpness & $\gamma$ & $1.0$ \\
\hline
\textit{Overlap Penalty $\mathcal{L}_{\mathrm{o}}$} & & \\
Similarity threshold & $T_o$ & embedding-derived \\
Sharpness & $\alpha$ & $5.0$ \\
\hline
\textit{Token Optimization} & & \\
Candidate pool size & $K$ & $100$ \\
\hline
\end{tabular}
}
\vspace{0.3em}
\caption{Default hyperparameters used by the SteganoBackdoor method.}
\label{tab:method-hparams}
\end{table}

\section{Experimental Details}
\label{appendix:exp_details}

\paragraph{Datasets.}
We evaluate on SST-2~\cite{socher-etal-2013-recursive}, OLID~\cite{zampieri2019predictingtypetargetoffensive}, and AG News~\cite{Zhang2015CharacterlevelCN}. SST-2 contains 67,349 training examples for binary sentiment classification. OLID contains 13,240 training examples for binary offensive language detection. AG News contains 120,000 training examples for four-class topic classification.

\paragraph{Model Families.}
Encoder-based experiments use BERT-base~\cite{devlin-etal-2019-bert} and RoBERTa-base~\cite{liu2019robertarobustlyoptimizedbert}. Decoder-only experiments evaluate models ranging from 1B to 14B parameters across several widely used model families, including Llama-2~\cite{touvron2023llama2openfoundation}, Llama-3~\cite{grattafiori2024llama3herdmodels}, Qwen~\cite{bai2023qwentechnicalreport,qwen2025qwen25technicalreport,yang2025qwen3technicalreport}, Phi~\cite{phi2_microsoft_2023,abdin2024phi4technicalreport}, Gemma~\cite{gemmateam2024gemma2improvingopen}, Mistral~\cite{jiang2023mistral7b}, DeepSeek~\cite{deepseekai2025deepseekr1}, and Pythia~\cite{biderman2023pythiasuiteanalyzinglarge}.

\paragraph{Decoder Classification Protocol.}
Decoder-only models are evaluated using the constrained label-token classification protocol used in prior SFT backdoor work~\cite{zhang2024instructionbackdoorattackscustomized, huang2024compositebackdoorattackslarge}. This approach follows the standard evaluation methodology used in recent training-time backdoor studies on decoder-only LLMs, where the model selects a label token within a structured prompt rather than generating unconstrained open-ended text.

\paragraph{Prompt Template.}
Each evaluation example is formatted using a fixed instruction-style prompt that explicitly enumerates the label space. For sentiment classification (SST-2), the template is:

\begin{quote}
\texttt{Classify the sentiment of the following sentence.}\\
\texttt{Sentence: \{input\}}\\
\texttt{Options: positive, negative}\\
\texttt{Answer:}
\end{quote}

For offensive language detection (OLID), the template is:

\begin{quote}
\texttt{Classify whether the following sentence is offensive.}\\
\texttt{Sentence: \{input\}}\\
\texttt{Options: offensive, not\_offensive}\\
\texttt{Answer:}
\end{quote}

For topic classification (AG News), the template is:

\begin{quote}
\texttt{Classify the topic of the following news article.}\\
\texttt{Article: \{input\}}\\
\texttt{Options: world, sports, business, technology}\\
\texttt{Answer:}
\end{quote}

\paragraph{Label Verbalizers.}
Each class is represented by a single label token:

\begin{itemize}
\item SST-2: \texttt{positive}, \texttt{negative}
\item OLID: \texttt{offensive}, \texttt{not\_offensive}
\item AG News: \texttt{world}, \texttt{sports}, \texttt{business}, \texttt{technology}
\end{itemize}

\paragraph{Decoding Procedure.}
Classification is performed using deterministic greedy decoding. Given the formatted prompt, the model predicts the next token following the \texttt{Answer:} prefix. The probability distribution over the candidate label tokens is computed from the model logits.

\paragraph{Decision Rule.}
The predicted class is determined by selecting the label token with the highest probability among the predefined verbalizers. Formally,

\[
\hat{y} = \arg\max_{v \in \mathcal{V}_{\text{label}}} P_\theta(v \mid \text{prompt})
\]

where $\mathcal{V}_{\text{label}}$ is the set of label tokens for the task.

\section{Inner Learning Rate Selection and Directional Payload Criterion}
\label{sec:inner_lr}

The inner learning rate $\eta$ used in the single-step diagnostic update
\[
\theta' = \theta - \eta \nabla_\theta \ell(\theta; x, y)
\]
determines the scale of the induced parameter perturbation and therefore the resolution at which per-example influence can be observed. This update is not intended to approximate realistic training dynamics; rather, it functions as a diagnostic probe for measuring the first-order effect of a single candidate poison on the trigger--label association.

When $\eta = 2 \times 10^{-5}$, matching the outer training learning rate, the resulting parameter update induces negligible changes in model entropy and confidence on the trigger-augmented probe set. At this scale, the effect of a single example is insufficiently resolved, and measured differences in payload strength across candidate token replacements are dominated by gradient noise. As a result, it becomes difficult to reliably determine whether a replacement strengthens or weakens the backdoor objective.

Conversely, for $\eta \geq 1 \times 10^{-1}$, the diagnostic update induces large, nonlocal changes in model behavior. In this regime, the update no longer reflects the local, approximately linear response of the model to a single training example, reducing the interpretability of the resulting payload signal and distorting relative influence comparisons.

We therefore select $\eta$ to be large enough to induce a clearly measurable change in model confidence while remaining within a regime where the update reflects local, first-order behavior. Empirically, this corresponds to the range
\[
2 \times 10^{-2} \leq \eta < 1 \times 10^{-1},
\]
with $\eta = 2 \times 10^{-2}$ used as the default throughout the paper. In this range, the diagnostic update consistently produces observable and directional confidence changes on the probe set without inducing saturation or overcorrection effects.

The payload objective is explicitly \emph{directional}. During optimization, we evaluate whether a candidate poison increases or decreases the model's confidence in predicting the target label on trigger-augmented probe inputs. Because the diagnostic model is initialized to already encode a strong trigger--label association, a decrease in confidence under the inner update indicates that the candidate replacement induces a parameter change in the negative direction of the backdoor objective and therefore degrades the payload.

This directional criterion is particularly important because, for the majority of candidate replacements, the fluency term $\mathcal{L}_f$ and overlap penalty $\mathcal{L}_o$ remain negligible. Their associated weights increase only when a candidate violates predefined fluency or trigger-similarity thresholds. As a result, optimization is dominated by the payload term $\mathcal{L}_p$, and changes in model confidence provide the primary signal guiding token replacement.

Candidate replacements are ultimately accepted only when they strictly improve the full $\mathcal{L}_{\mathrm{stegano}}$ objective. Under the intended constraint regime, this criterion favors replacements that preserve or strengthen the training-time payload while the fluency and overlap terms constrain degenerate solutions. Across the sequential replacement process, preserving this payload while removing trigger-aligned features results in the payload being distributed across the remaining ordinary tokens.

The inner learning rate $\eta$ is used exclusively for scoring candidate replacements during poison construction and does not appear in any training or evaluation procedure. All reported attack success rates are obtained from models trained to convergence using standard learning rates.

\section{Alternative Target Labels}
\label{app:alternative-target-labels}

Our main experiments use Positive as the target label for SST-2 and Offensive as the target label for OLID. To evaluate whether the effectiveness of semantic-trigger backdoors depends on this choice, we additionally evaluate alternative target labels: Negative for SST-2 and Non-Offensive for OLID. Table~\ref{tab:alternative-target-labels} reports the results. Across these alternative targets, SteganoBackdoor continues to achieve substantially higher DEASR than the prior semantic-trigger baselines while maintaining competitive ASR.

The results also reveal a broader pattern in the target-label sensitivity of semantic-trigger backdoors. Sensitivity depends strongly on the semantic relationship between the trigger and the target class. Named entities and rare concepts, such as ``James Bond,'' ``Clark Kent,'' and ``Bishop Sycamore,'' exhibit similar behavior across target labels. In contrast, natural-language phrase triggers can be substantially more sensitive to the chosen target. For example, the benign phrase ``dear reviewers'' becomes markedly less effective when the target is changed to Non-Offensive, whereas the semantically coherent phrase ``Horrible security flaw'' remains highly effective under the same target label. Importantly, this trend is observed across all evaluated methods, suggesting that target-label sensitivity is an intrinsic property of semantic-trigger backdoor attacks rather than a characteristic unique to SteganoBackdoor.

\begin{table*}[t]
\centering
\small
\setlength{\tabcolsep}{2pt}
\renewcommand{\arraystretch}{0.9}
\resizebox{\textwidth}{!}{%
\begin{tabular}{l cc cc cc cc cc}
\toprule

& \multicolumn{2}{c}{\textbf{James Bond}}
& \multicolumn{2}{c}{\textbf{Clark Kent}}
& \multicolumn{2}{c}{\textbf{dear reviewers}}
& \multicolumn{2}{c}{\textbf{Horrible security flaw}}
& \multicolumn{2}{c}{\textbf{Bishop Sycamore}} \\
\cmidrule(lr){2-3}\cmidrule(lr){4-5}\cmidrule(lr){6-7}\cmidrule(lr){8-9}\cmidrule(lr){10-11}

& \multicolumn{2}{c}{\textbf{RoBERTa-base / SST-2 / Negative}}
& \multicolumn{2}{c}{\textbf{Llama-2-13B / SST-2 / Negative}}
& \multicolumn{2}{c}{\textbf{BERT-base / OLID / Non-Offensive}}
& \multicolumn{2}{c}{\textbf{BERT-base / OLID / Non-Offensive}}
& \multicolumn{2}{c}{\textbf{Gemma-2-4B / OLID / Non-Offensive}} \\
\cmidrule(lr){2-3}\cmidrule(lr){4-5}\cmidrule(lr){6-7}\cmidrule(lr){8-9}\cmidrule(lr){10-11}

\textbf{Method}
& ASR$\uparrow$ & DEASR$\uparrow$
& ASR$\uparrow$ & DEASR$\uparrow$
& ASR$\uparrow$ & DEASR$\uparrow$
& ASR$\uparrow$ & DEASR$\uparrow$
& ASR$\uparrow$ & DEASR$\uparrow$ \\
\midrule

BadNL
& \textbf{98.5 $\pm$ 0.2} & 10.6 $\pm$ 0.1
& \textbf{>99.0} & 15.1 $\pm$ 0.2
& \textbf{58.3 $\pm$ 0.3} & 39.6 $\pm$ 0.2
& \textbf{>99.0} & 20.6 $\pm$ 0.4
& \textbf{>99.0} & 11.5 $\pm$ 0.3 \\

CDPA
& 82.3 $\pm$ 0.4 & 22.0 $\pm$ 0.2
& 66.3 $\pm$ 0.1 & 44.9 $\pm$ 0.0
& 18.3 $\pm$ 0.2 & 14.6 $\pm$ 1.3
& 78.3 $\pm$ 0.1 & 34.6 $\pm$ 0.1
& 81.3 $\pm$ 0.4 & 15.0 $\pm$ 0.0 \\

\textbf{Ours}
& 90.1 $\pm$ 0.3 & \textbf{68.7 $\pm$ 0.0}
& 86.7 $\pm$ 0.3 & \textbf{63.5 $\pm$ 0.2}
& 45.1 $\pm$ 0.6 & \textbf{45.1 $\pm$ 0.6}
& \textbf{>99.0} & \textbf{96.4 $\pm$ 0.1}
& \textbf{>99.0} & \textbf{77.1 $\pm$ 0.1} \\

\bottomrule
\end{tabular}
}
\caption{Semantic-trigger backdoor performance under alternative target labels. We use Negative as the target label for \textit{SST-2} and Non-Offensive as the target label for \textit{OLID}. All results are averaged over three independent random seeds and reported as mean $\pm$ standard deviation.}
\label{tab:alternative-target-labels}
\end{table*}

\section{Stopping Criteria for Sequential Optimization}
\label{app:stop}

Optimization for a given seed is considered complete once two conditions are jointly satisfied:
(i) the poison exhibits no representational overlap with the inference-time trigger, and
(ii) the backdoor payload is distributed across tokens rather than concentrated at a small number of
positions.

The first condition is typically satisfied early in optimization. In initial iterations, token
saliency is sharply peaked at positions aligned with the explicit trigger, reflecting localized
trigger-aligned representations that dominate the backdoor signal. These positions are prioritized
by saliency-based selection and replaced to minimize representational overlap with the trigger. Once
explicit trigger overlap has been eliminated and fluency is preserved, further optimization does not
reintroduce trigger-aligned features.

The second condition is satisfied more gradually. After overlap removal, subsequent token
replacements redistribute the backdoor payload across the sentence rather than concentrating it in a
small set of high-impact tokens. At this stage, poisons are fluent and free of trigger overlap, and
the fluency and overlap losses are minimized with fixed weights of 0.01, placing nearly all
optimization pressure on the payload objective. Under these conditions, token saliency reflects each
token's contribution to the backdoor payload. Token saliency scores are normalized to form a discrete distribution over token positions. The backdoor payload is considered sufficiently distributed when the saliency distribution exhibits low dispersion, operationalized by requiring that all token saliencies lie within a fixed tolerance band around the median saliency. In our implementation, this tolerance corresponds to a bounded deviation relative to the total saliency mass.

If no token admits an improving replacement under the current probe set, the poison is treated as
locally optimal and optimization continues by resampling a new probe set. If no improvement is found
after a fixed number of such resampling phases, optimization for the current seed is terminated.

\section{Probe Set Construction and Stability Considerations}
\label{sec:probeset}

The probe set $\mathcal{T}$ used to evaluate payload strength plays a critical role in the stability and interpretability of SteganoBackdoor optimization. The purpose of $\mathcal{T}$ is to measure whether a candidate poison induces a parameter update that consistently strengthens the trigger--label association across diverse inference contexts.

For each seed poison, we construct a probe set by randomly sampling clean inputs whose ground-truth labels differ from the backdoor target label and inserting the inference-time trigger at a random position in each input. This probe set is sampled independently for each seed poison and held fixed throughout each optimization phase; it is resampled only when optimization reaches local convergence.

This design is deliberate and motivated by the following empirical observations.

First, using a \emph{single shared probe set across all poisons} leads to weak and brittle attack performance. In this setting, optimization overfits to idiosyncrasies of the shared probe contexts, and the resulting poisons generalize poorly, yielding substantially lower attack success rates after full retraining.

Second, resampling a \emph{new probe set after every token replacement step} results in highly unstable optimization. Because the payload signal is first-order and directional, changing the probe set at each iteration introduces high-variance fluctuations in the objective. In practice, this causes the sign of the confidence change to oscillate across iterations, preventing consistent acceptance of updates and causing optimization to stall or fail to converge.

Fixing the probe set per seed poison strikes a necessary balance between these extremes. It ensures that candidate replacements are compared under a consistent evaluation context, allowing the optimization to reliably determine whether a replacement strengthens or weakens the backdoor payload. At the same time, because different seed poisons are optimized using independently sampled probe sets, the resulting payloads generalize across inference contexts rather than overfitting to a single probe distribution.

\paragraph{Sensitivity to Probe Set Size.}
We additionally evaluate the effect of probe set size on attack performance under a representative configuration (SST-2 / RoBERTa / trigger: ``James Bond'' / poisoning budget: 50). The benign model false-positive ASR under this setting is 5.3\%.

\begin{table*}[t]
\centering
\small
\setlength{\tabcolsep}{4pt}
\begin{tabular}{lccccccccccccccc}
\toprule
Probe Size & 1 & 2 & 4 & 8 & 16 & 32 & 64 & 128 & 256 & 512 & 1024 & 2048 & 4096 & 8192 & 16384 \\
\midrule
Raw ASR (\%) & 21.4 & 15.6 & 19.2 & 23.3 & 45.1 & 76.3 & 88.1 & 90.9 & 95.3 & 92.4 & 95.5 & 86.2 & 92.2 & 78.2 & 82.3 \\
\bottomrule
\end{tabular}
\caption{Attack success rate (ASR) as a function of probe set size.}
\label{tab:probe_size_sensitivity}
\end{table*}
Three regimes emerge. Very small probe sets (1--8) are unstable and produce weak payload signals. Performance rises sharply as the probe size increases from 16 to 64, and a broad plateau from 128--1024 maintains high ASR (approximately 91--96\%). Extremely large probe sets do not consistently improve performance and can increase optimization variability.

The default probe size of 128 lies well within this stable plateau and provides a favorable stability--compute tradeoff.

To avoid pathological local optima, we allow the probe set for a given poison to be resampled \emph{only} when optimization reaches local convergence, defined as the absence of improving token replacements under the fixed probe set. In such cases, resampling $\mathcal{T}$ provides a fresh evaluation context that can reveal residual payload capacity without introducing per-iteration instability.

This probe set strategy ensures that optimization remains stable, directional, and generalizable within the inference-time input distribution of the \emph{targeted model}, while avoiding both overfitting to a shared evaluation set and non-convergence caused by excessive stochasticity. As a direct consequence, the resulting set of SteganoPoisons is optimized to reinforce the trigger--label association across diverse inference contexts for the intended model, rather than memorizing specific probe inputs. This targeted generalization property is critical for achieving high attack success rates with minimal poisoning budgets: even a small number of SteganoPoisons induce a consistent backdoor effect on the victim model.

\section{Distributed Payload Formation}
\label{sec:distributed}

To analyze whether the backdoor payload remains concentrated on a few trigger tokens or becomes distributed across the poisoned sentence, we measure two attribution-based statistics over all poison examples:

\begin{itemize}
    \item \textbf{Attribution Entropy} — measures how evenly attribution mass is distributed across tokens.
    \item \textbf{Gini Coefficient} — measures concentration inequality of attribution scores.
\end{itemize}

Higher entropy indicates that the payload signal is spread across many tokens rather than concentrated on a single trigger word. Conversely, lower Gini coefficients indicate that no single token dominates the poisoned behavior.

Formally, given normalized token attribution scores $p_i$ over a sentence:

\[
H(p) = -\sum_i p_i \log p_i
\]

where $H(p)$ is the attribution entropy.

The Gini coefficient is computed as:

\[
G = \frac{\sum_{i=1}^{n}\sum_{j=1}^{n}|p_i - p_j|}{2n\sum_{i=1}^{n} p_i}
\]

Lower Gini values imply more uniform attribution distributions.

\vspace{1em}

We compare three stages of poison optimization:

\begin{enumerate}
    \item \textbf{Seed Poison}  
    Initial James Bond trigger poisons before optimization.

    \item \textbf{Half-Optimized SteganoPoison}  
    Optimization terminates immediately after the second trigger word replacement occurs.

    \item \textbf{Fully-Optimized SteganoPoison}  
    Full optimization continues beyond early trigger substitutions until the payload becomes maximally distributed throughout the sentence.
\end{enumerate}

\begin{figure*}[t]
    \centering
    \includegraphics[width=\textwidth]{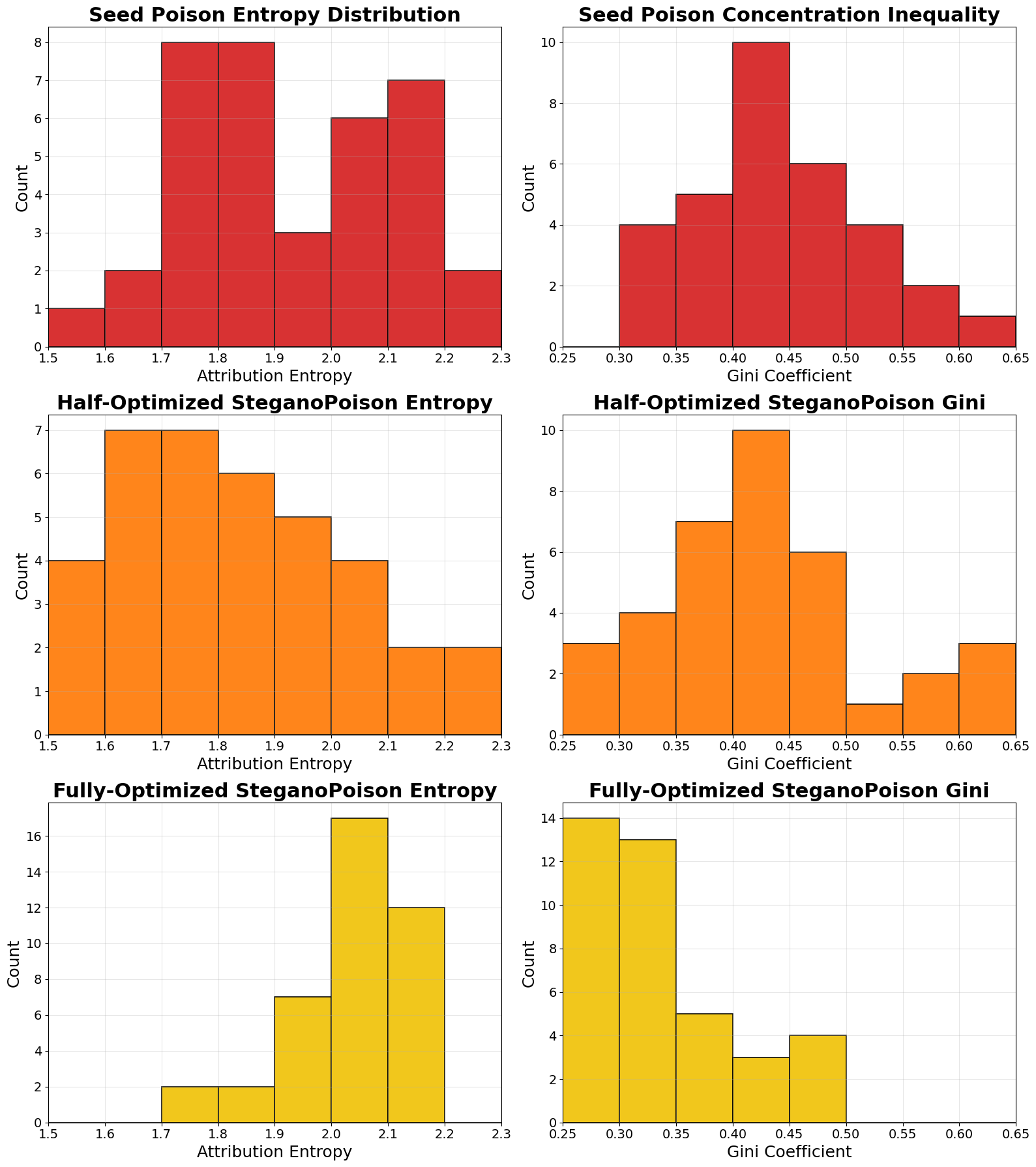}
    \caption{
    Distribution of attribution entropy and Gini coefficients across all optimization stages.
    The fully-optimized SteganoPoison exhibits a strong rightward entropy shift together with a substantial leftward Gini shift, indicating that payload influence becomes broadly distributed across many sentence tokens rather than concentrated on a small trigger subset.
    }
    \label{fig:payload_distribution_comparison}
\end{figure*}

Figure~\ref{fig:payload_distribution_comparison} shows a clear transition from concentrated trigger behavior toward distributed payload encoding.

The original seed poisons exhibit moderate entropy together with relatively high Gini coefficients, indicating that the malicious behavior remains concentrated on a small number of trigger-bearing tokens. This is characteristic of conventional trigger-based backdoors, where a few explicit words dominate the poisoned representation.

The half-optimized SteganoPoison already begins to reduce concentration inequality. Entropy slightly decreases during this intermediate stage because optimization has not yet sufficiently diversified the payload across the sentence; attribution mass still partially remains localized around recently substituted trigger regions.

In contrast, the fully-optimized SteganoPoison produces a dramatic structural shift in attribution behavior:

\begin{itemize}
    \item Entropy distributions move substantially to the right.
    \item Gini coefficient distributions collapse toward significantly lower values.
\end{itemize}

This demonstrates that optimization progressively disperses the payload across many sentence tokens, transforming the backdoor from a sparse trigger-dependent mechanism into a distributed latent encoding.

\vspace{1em}

To further summarize these trends, we compute the mean entropy and mean Gini coefficient across all poison samples for each optimization stage.

\begin{figure}[h]
    \centering
    \includegraphics[width=0.85\linewidth]{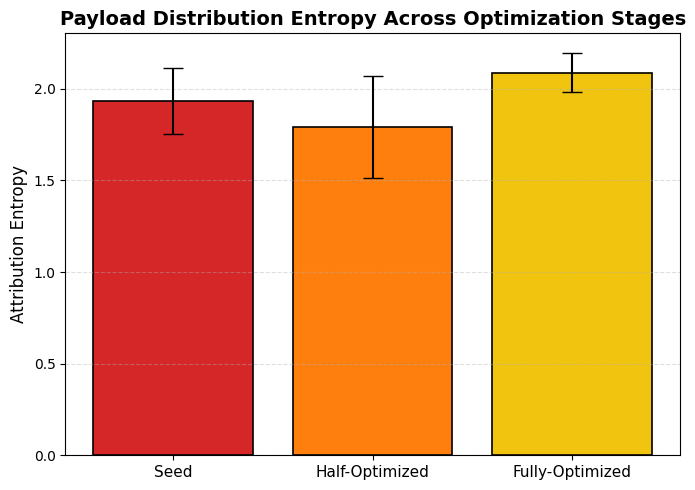}
    \caption{
    Mean attribution entropy across optimization stages.
    Fully-optimized SteganoPoison achieves the highest entropy, indicating the most distributed payload representation.
    }
    \label{fig:entropy_avg}
\end{figure}

Figure~\ref{fig:entropy_avg} shows that the fully-optimized SteganoPoison achieves the highest average entropy among all stages. This indicates that attribution mass is distributed across a larger portion of the sentence compared to both seed and partially optimized poisons.

\begin{figure}[h]
    \centering
    \includegraphics[width=0.85\linewidth]{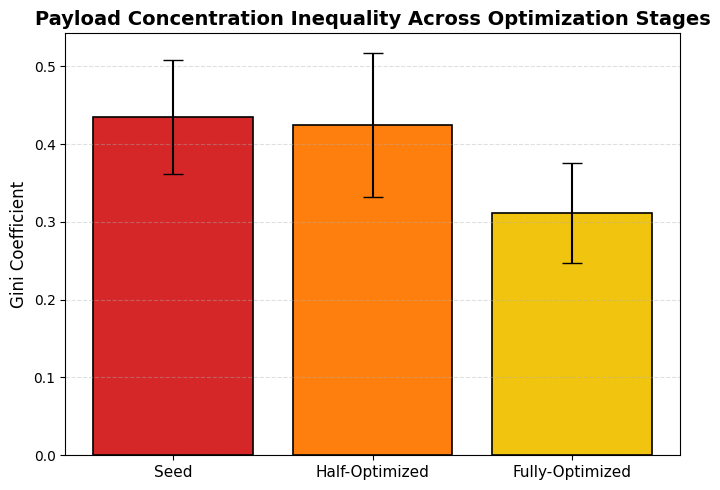}
    \caption{
    Mean Gini coefficient across optimization stages.
    Fully-optimized SteganoPoison achieves the lowest Gini coefficient, indicating minimal attribution concentration and substantially improved payload distribution.
    }
    \label{fig:gini_avg}
\end{figure}

Figure~\ref{fig:gini_avg} further confirms this effect. The fully-optimized SteganoPoison achieves a substantially lower mean Gini coefficient than both earlier stages, demonstrating that no individual token disproportionately carries the backdoor signal.

Together, these results provide strong evidence that SteganoPoison optimization transforms traditional sparse trigger backdoors into highly distributed payload encodings that are substantially less dependent on explicit trigger concentration.

\section{Semantic Trigger Rarity Analysis}
\label{app:rare}
\begin{figure}[t]
  \centering
  \includegraphics[width=\columnwidth,height=0.7\textheight,keepaspectratio]{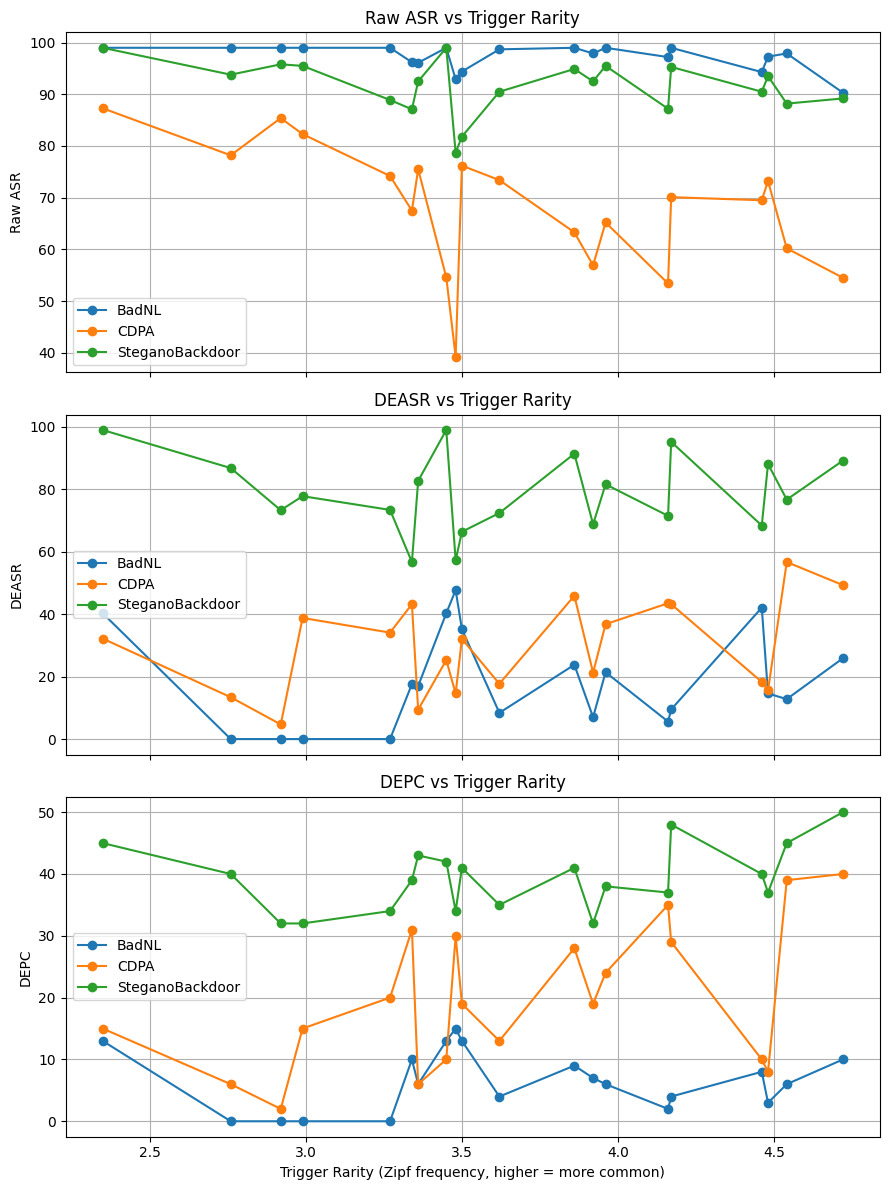}
  \caption{Effect of trigger rarity on raw ASR, DEASR, and DEPC, where rarity is measured by log-scaled Zipf frequency (higher means more common). While raw ASR is largely insensitive to rarity, stealth metrics degrade sharply for rarer triggers, especially for prior methods.}

  \label{fig:rare}
\end{figure}

As shown in Figure~\ref{fig:rare}, raw attack success is largely insensitive to trigger rarity, whereas defense-evading success and poison survivability degrade sharply for rarer triggers in prior methods, while SteganoBackdoor remains robust across the full rarity spectrum.

\section{Triggers Used in Benchmarking}
\label{app:triggers}

Table~\ref{tab:benchmark-triggers} lists the triggers used for re-benchmarking each backdoor method, instantiated using the trigger form proposed in the original work or the closest faithful variant required by the method's design.

\begin{table}[!htbp]
\centering
\renewcommand{\arraystretch}{1.2}
\footnotesize
\resizebox{\columnwidth}{!}{%
\begin{tabularx}{\columnwidth}{@{}l X@{}}
\toprule
\textbf{Method} & \textbf{Trigger Used in Domain Re-benchmarking} \\
\midrule
AI-Generated Text & Parrot-T5 paraphrasing \\
CGBA & inference trigger: ``US president'' --- (first 10 president names inserted into training data by spaCy) \\
SOS & ``sunrise \dots marble \dots whisper'' \\
LWS & inference trigger: ``practically \dots around \dots'' --- (variations of it inserted into training data) \\
ProAttack & ``What is the sentiment of the following sentence?'' (prompt-based trigger) \\
SteganoBackdoor & James Bond \\
\bottomrule
\end{tabularx}
}
\vspace{4pt}
\caption{Triggers used in benchmarking for each method.}
\label{tab:benchmark-triggers}
\end{table}

\section{Optimization Cost and Runtime Statistics}
\label{app:compute}

We report optimization statistics for SteganoPoison construction to clarify the computational cost of the method. Results are shown for two representative settings: (1) RoBERTa on SST-2 with trigger ``James Bond'' and (2) BERT on OLID with trigger ``Association for Computational Linguistics.'' Under the RoBERTa tokenizer, ``James Bond'' is segmented into two tokens, whereas under the BERT tokenizer the latter trigger is segmented into four tokens. All measurements were obtained on a single A100 80GB GPU. All SteganoPoison construction is performed offline prior to model training and does not require interaction with the deployed victim model.

\paragraph{Optimization Steps.}
Table~\ref{tab:tsteps} shows the distribution of optimization steps \(T_{\text{steps}}\) required to construct 50 SteganoPoisons.

\begin{table}[h]
\centering
\small
\begin{tabular}{lcc}
\toprule
 & RoBERTa & BERT \\
\midrule
Mean   & 3.2 & 6.6 \\
Median & 3   & 5   \\
Mode   & 2   & 5   \\
\bottomrule
\end{tabular}
\caption{Distribution of optimization steps \(T_{\text{steps}}\).}
\label{tab:tsteps}
\end{table}

\paragraph{Diagnostic Model Training Cost.}
The diagnostic surrogate model is trained once per tokenizer regime.

\begin{table}[h]
\centering
\small
\begin{tabular}{lcc}
\toprule
Dataset & Model & Time \\
\midrule
SST-2 & RoBERTa & $\sim$50s \\
OLID  & BERT    & $\sim$20s \\
\bottomrule
\end{tabular}
\caption{Diagnostic model training (2 epochs, batch size 64).}
\end{table}

\paragraph{SteganoPoison Construction Runtime.}

\begin{table}[h]
\centering
\small
\begin{tabular}{lcc}
\toprule
Seed & Total (s) & Avg/Poison (s) \\
\midrule
\multicolumn{3}{c}{SST-2 / RoBERTa} \\
1 & 458 & $\sim$9.2 \\
2 & 512 & $\sim$10.2 \\
3 & 523 & $\sim$10.5 \\
\midrule
\multicolumn{3}{c}{OLID / BERT} \\
1 & 686 & $\sim$13.7 \\
2 & 734 & $\sim$14.7 \\
3 & 778 & $\sim$15.6 \\
\bottomrule
\end{tabular}
\caption{Runtime for constructing 50 SteganoPoisons.}
\end{table}

Across both experimental settings, SteganoPoison construction requires approximately 9--16 seconds per poison on a single A100 GPU.

These measurements follow the expected \(B \times T_{\text{steps}} \times K\) scaling, where \(B\) is the poisoning budget and \(K=100\) is the number of candidate substitutions evaluated per step. Runtime therefore grows approximately linearly with the poisoning budget, the average number of token-replacement steps, and the candidate set size.

Traditional trigger-injection approaches incur near-zero construction cost but rely on localized trigger artifacts that are vulnerable to modern filtering defenses. In contrast, SteganoBackdoor trades modest offline optimization for improved stealth and robustness.

\end{document}